\documentclass[%
 reprint,
 amsmath,amssymb,
 aps,
 prx,
]{revtex4-2}

\usepackage{graphicx}
\usepackage{svg}
\usepackage{stmaryrd}
\usepackage{dcolumn}
\usepackage{bm}
\usepackage[unicode,pdfusetitle,colorlinks,allcolors=blue]{hyperref}
\usepackage{accents}
\usepackage{xcolor}
\usepackage{needspace}

\usepackage{circuitikz}
\usetikzlibrary{calc, positioning, arrows.meta, fit, backgrounds, shapes.geometric, shapes.misc, decorations.pathmorphing}
\usepackage{algorithm}
\usepackage{algpseudocode}
\usepackage{float}
\usepackage{subcaption}
\usepackage{enumitem}
\usepackage{listings}
\usepackage{quantikz}
\usepackage{booktabs}
\usepackage{comment}

\definecolor{extGold}{HTML}{F5D64C}
\definecolor{extSolar}{HTML}{FFF1C8}
\definecolor{extYellow}{HTML}{FFB400}
\definecolor{extOrange}{HTML}{FF8400}
\definecolor{extCopper}{HTML}{903001}
\definecolor{extFuchsia}{HTML}{4E012A}
\definecolor{extBrown}{HTML}{1C0101}

\newcommand{\btheta}{\boldsymbol{\theta}}

\newcommand{\R}{\mathbb{R}}                           
\newcommand{\Ex}{\mathbb{E}}                         

\newcommand{\dd}{\mathrm{d}}

\newcommand{\Cov}{\mathrm{Cov}}

\newcommand{\pket}[1]{|#1)}
\newcommand{\pbra}[1]{(#1|}

\newcommand{\gname}[1]{\ensuremath{\mathsf{#1}}}          
\newcommand{\Id}{\ensuremath{\mathbb{I}}}                 
\newcommand{\graph}{\ensuremath{\mathcal{G}}}             
\newcommand{\pbit}{\ensuremath{\mathsf{pbit}}}
\newcommand{\pdit}{\ensuremath{\mathsf{pdit}}}
\newcommand{\pmode}{\ensuremath{\mathsf{pmode}}}

\newcommand{\creal}{\ensuremath{\mathsf{real}}}

\begin{document}


\title{A Framework for Stochastic Differentiable Programming}
\author{Guillaume Verdon} \affiliation{Extropic Corporation, San Francisco, California 94111, USA}
 \author{Leo Tyrpak} \affiliation{Extropic Corporation, San Francisco, California 94111, USA}
 \author{Owen Lockwood} \affiliation{Extropic Corporation, San Francisco, California 94111, USA}
 \author{Seth Morton} \affiliation{Extropic Corporation, San Francisco, California 94111, USA}
\author{Alexander Neagoe} \affiliation{Extropic Corporation, San Francisco, California 94111, USA}
\author{Anton Sugolov} \affiliation{Extropic Corporation, San Francisco, California 94111, USA}
\author{Ian MacCormack} \affiliation{Extropic Corporation, San Francisco, California 94111, USA}
\author{Mirko Amico} \affiliation{Extropic Corporation, San Francisco, California 94111, USA}


\date{\today}
\begin{abstract}
We introduce Parametrized Stochastic Circuits (PSCs), a gate-based intermediate representation for programmable stochastic dynamics in which typed local stochastic kernels with tunable parameters compose over explicit binary, categorical, and continuous wires, and \texttt{torx}, an open-source JAX framework for constructing, executing, and differentiating them. PSC's data types and stochastic kernels are chosen to align closely with the native operations exposed by emerging probabilistic hardware. In this way, stochastic algorithms can be designed directly in terms of the operations the hardware executes natively, so that the energy advantage arising at this level is not lost on mappings that introduce substantial decomposition, communication, or control overhead. We demonstrate the framework on a variety of example applications such as random walks on graphs, discrete diffusion, stochastic graph networks, jump diffusion and Ising sampling. We also report a hardware experiment in which probabilistic bits on the X0 subthreshold CMOS test chip, hosted by the XTR-0 desktop platform, provide physical randomness for Metropolis-Hastings and importance-sampling estimators, yielding estimates consistent with a software pseudorandom baseline.
\end{abstract}

\maketitle

\section{Introduction}\label{sec:intro}

Many modern algorithms are built from parametrized stochastic maps. Diffusion models learn denoising transitions that turn noise into data~\cite{ho2020denoising}, Monte Carlo methods chain transition kernels to estimate high-dimensional expectations~\cite{neal2011mcmc}, probabilistic circuits represent distributions through compositions of tractable factors~\cite{choi2020probabilistic,vergari2021compositional}, and regime-switching models couple a latent discrete process to a continuous observable~\cite{hamilton1989new}. These methods share a compositional structure: local probabilistic operations are assembled into a global computation, and their parameters may be fixed or optimized against scalar objectives defined on samples or output distributions. A logical composition, however, does not by itself specify how its operations are realized. In conventional software this distinction is largely an implementation detail but on specialized stochastic hardware, it can determine the feasibility and cost of execution.

Emerging thermodynamic and probabilistic devices implement physical stochastic primitives at low energy cost. Some devices provide physical random sources that can be consumed by otherwise digital algorithms, such as probabilistic bits built from magnetic tunnel junctions~\cite{chowdhury2023full,zhang2025automatic} or subthreshold CMOS circuits~\cite{freitas2026taming}, whereas other devices implement families of stochastic kernels directly through their physical dynamics, such as Ising machines executing Gibbs sampling natively in silicon~\cite{jelinvcivc2025efficient} and superconducting circuits realizing continuous-variable computations at thermodynamic equilibrium~\cite{lockwood2026blueprint}. These devices differ in their supported variable types, transition families, connectivity, and control interfaces. Any advantage depends on the total cost of realizing the logical operations on the device. Lowering a high-level stochastic map through a complex sequence of physical primitives, communication steps, or digital computations may eliminate the potential advantage of the substrate. This motivates a hardware-near intermediate representation that specifies the logical composition of stochastic kernels while keeping the structure and cost of their physical realization visible.

Existing software frameworks for stochastic computing focus mainly on providing composable primitives at a high level of abstraction. Stochastic computation graphs formalize computations as directed graphs of deterministic operations and conditional random variables and give general rules for constructing unbiased gradient estimators~\cite{schulman2015stochastic}. Pyro~\cite{bingham2019pyro} and NumPyro~\cite{phan2019numpyro} provide composable modeling and inference primitives on differentiable array frameworks, while TensorFlow Probability exposes composable Markov-transition kernels for constructing and executing MCMC algorithms~\cite{lao2020tfpmcmc}. The frameworks establish that stochastic maps can be represented, composed and executed in multiple ways. However, their primitives are software constructs, distribution objects, tensor operations, and inference routines, with no direct correspondence to the typed variables and local stochastic kernels a device natively provides. Gate-and-circuit abstractions in quantum software frameworks such as TensorFlow-Quantum~\cite{broughton2020tensorflowquantum} and PennyLane~\cite{bergholm2018pennylane} faced the same constraint, where the scarce resource in noisy quantum devices is coherence rather than energy, and settled on the circuit as the lowest level of abstraction that remains intelligible and differentiable while keeping the mapping to hardware simple. However, an analogous solution for stochastic computation has not yet appeared.

In this work, we introduce Parametrized Stochastic Circuits (PSCs), a gate-based model of programmable stochastic dynamics which serves as such an intermediate representation, together with \texttt{torx}~\cite{torxlib}, an open-source JAX~\cite{jax2018github,kidger2021equinox} framework for constructing, executing and differentiating them. A PSC is an ordered sequence of layers of stochastic gates acting on binary, categorical, and continuous wires, where each gate is a local conditional transition rule with tunable settings. Gates may be finite transition matrices, Markov kernels on continuous spaces, or finite-time operators obtained by exponentiating local rate models. The primitives of a PSC, typed variables and local stochastic kernels with explicit supports are close to the components a device natively provides while leaving some freedom to lower the logical description given by a PSC to match exactly the hardware constraints. All stochastic effects are localized to the gates, so one can fix the program semantics by defining a logical PSC and then choose a variety of execution methods. Exact execution propagates the full output distribution when the state space allows it. Sampled execution provides trajectories drawn from the circuit and is compatible with software and hardware randomness providers, which can be exchanged without modifying the logical PSC. In the case of devices implementing kernels via their physical dynamics, one can compile the logical PSC to a hardware-compatible representation, as described in the \texttt{thermalizers} framework of the companion manuscript~\cite{amico2026thermalizing}.

We demonstrate the framework on various example applications, from purely binary circuits to hybrid discrete-continuous ones, including random walks on graphs, Ising sampling, discrete diffusion, stochastic graph networks, and jump diffusion. We also report a hardware experiment in which calibrated probabilistic bits (pbits) on the X0 subthreshold CMOS test chip, hosted by the XTR-0 desktop platform, supply Bernoulli randomness to Metropolis-Hastings and importance-sampling estimators, with estimates consistent with a software pseudorandom baseline. The logical PSC is unchanged between the two runs, and the only difference between them is the randomness provider. 

The remainder of the manuscript is organized as follows. Section~\ref{sec:background} defines the computational primitives, stochastic kernels and their compositions. Section~\ref{sec:stoc_circs} defines PSCs, the elementary gate library, product-formula approximations, and the gradient estimation rules. Section~\ref{sec:torx-impl} describes the \texttt{torx} framework along with its software and hardware backends. Section~\ref{sec:examples} demonstrates the framework on the examples mentioned above, including the X0 hardware experiment, and the conclusions follow in Section~\ref{sec:discussion}. Appendices expand on details left out of the main text.

\section{Background and preliminaries}\label{sec:background}

This Section introduces the basic concepts which we build upon. We proceed at different levels of granularity, leading up to the gate-based circuit model of PSCs of Section~\ref{sec:stoc_circs}. In Section~\ref{sec:notation} we define the fundamental units of stochastic computation that we work with. Section~\ref{sec:kernels} defines stochastic kernels as the basic operations on data, the maps that transform states. We recall the definition of a Markov chain, a sequence of stochastic kernels, in Section~\ref{sec:markov}. Finally, composition of stochastic kernels into Directed Acyclic Graphs (DAGs) is covered in Section~\ref{sec:dfg}.

\subsection{Primitives and notation}
\label{sec:notation}

\begin{figure}[h]
    \centering
    \includegraphics[width=\columnwidth]{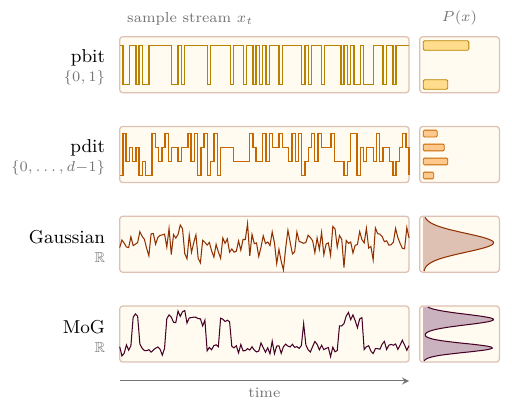}
    \caption{The data primitives of Table~\ref{tab:wire-types}. Each row shows an example of a sample stream $x_t$ together with the distribution $P(x)$ from which the samples are drawn. A pbit carries a distribution over two states, a pdit a distribution over \(d\) states, and a pmode a probability density over \(\mathbb{R}\); the Gaussian and mixture-of-Gaussians rows are two examples of distributions carried by a pmode wire.}
    \label{fig:primitives-family}
\end{figure}

We consider three data types, distinguished by their state space, summarized in Table~\ref{tab:wire-types} and illustrated in Figure~\ref{fig:primitives-family}. Throughout, we refer to a variable carrying one of these data types as
a \emph{wire}, in analogy with circuit diagrams; the circuits acting
on wires are defined in Section~\ref{sec:stoc_circs}.

\begin{table}[h]
\centering
\caption{Data primitives and their state spaces.}
\label{tab:wire-types}
\begin{tabular}{ll}
\toprule
Data type & State space \(\mathcal{X}\) \\
\midrule
\pbit{} (\(2\) states) & \(\{0,1\}\) \\
\pdit{} (\(d\) states) & \(\{0,\ldots,d-1\}\) \\
\pmode{} (\(N\) dimensions) & \(\R^{N}\) \\
\bottomrule
\end{tabular}
\end{table}

A \emph{pbit} is the binary case: its state space is \(\mathcal{X}=\{0,1\}\) and its state is a length-\(2\) probability vector \(\rho=(\rho_0,\rho_1)\in\Delta_1\), where \(\Delta_{d-1}=\bigl\{\rho\in\R^{d}:\rho_a\geq 0,\ \textstyle\sum_a\rho_a=1\bigr\}\) denotes the probability simplex on \(d\) states.
 
A \emph{pdit} generalizes this to \(d\) states~\cite{duffee2025pdit}: its state space is \(\mathcal{X}=\{0,\ldots,d-1\}\) and its state is a length-\(d\) probability vector \(\rho\in\Delta_{d-1}\).
 
A \emph{pmode} is a continuous wire with state space \(\mathcal{X}=\R^{N}\); its state is a probability density \(\rho(x)\geq 0\) satisfying \(\int_{\R^{N}}\rho(x)\,dx=1\).  The scalar case \(N=1\) is assumed
when \(N\) is left unspecified. While a pmode can carry an arbitrary continuous distribution, the elementary pmode gates of Section~\ref{sec:gates} act on Gaussian densities for tractability; non-Gaussian generator-derived gates (e.g.~\(\gname{PGBM}\), \(\gname{PCIR}\)) are catalogued in Appendix~\ref{app:pmode}.

Combining two state spaces \(\mathcal{X}_1\) and \(\mathcal{X}_2\) gives a joint state space \(\mathcal{X}_1\times\mathcal{X}_2\) whose elements are pairs \((a_1,a_2)\). A register of \(n\) pbits, for example, has state space \(\{0,1\}^n\) and state \(\rho\in\Delta_{2^n-1}\).

We adopt Dirac-style notation for deterministic states:
\(\pket{a_1\ldots a_n}\) denotes the distribution on \(\{0,1\}^n\) that
assigns probability \(1\) to the string \((a_1,\ldots,a_n)\),
\begin{equation}
  \pket{a_1\ldots a_n} \;=\; \delta_{(\,\cdot\,),(a_1,\ldots,a_n)}.
\end{equation}
Every distribution on \(\{0,1\}^n\) is then a convex combination
\(\rho=\sum_{a_1,\ldots,a_n}\lambda_{a_1\ldots a_n}\pket{a_1\ldots a_n}\).
This notation is inspired by quantum computation for convenience and used
only where it aids clarity.

With the states fixed, the next step is to define the maps that act on them.

\subsection{Stochastic Kernels}
\label{sec:kernels}

\begin{figure}[t]
\centering
\begin{tikzpicture}[
  >=stealth,
  font=\small,
  pwire/.style={thick, black!80},
  gate/.style={draw, thick, fill=extGold!18, draw=extCopper!75!black,
               rounded corners=2pt, inner sep=2pt,
               minimum width=0.55cm, minimum height=0.45cm},
  prep/.style={isosceles triangle, isosceles triangle apex angle=70,
               draw=black!80, fill=black!80, minimum width=0.0cm, minimum height=0.2cm,
               inner sep=0pt, anchor=apex},
  ground/.style={inner sep=0pt, minimum size=0pt},
  plabel/.style={font=\scriptsize, anchor=south, text=black!65, align=center},
  ilabel/.style={font=\scriptsize, anchor=center, text=black!60},
  wend/.style={font=\scriptsize, anchor=west, text=black!75},
  wbeg/.style={font=\scriptsize, anchor=east, text=black!75},
]
\begin{scope}[xshift=0cm, yshift=0cm]
  \draw[pwire] (-0.8, 0) -- (0.8, 0);
  \node[gate] at (0, 0) {\(K\)};
  \node[wbeg] at (-0.8, 0) {\(x\)};
  \node[wend] at ( 0.8, 0) {\(y\)};
  \node[plabel] at (0, 0.65) {(a) kernel};
\end{scope}
\begin{scope}[xshift=3.0cm, yshift=0cm]
  \draw[pwire] (-0.8, 0) -- (0.8, 0);
  \node[ilabel] at (0, -0.3) {\(\mathrm{Id}\)};
  \node[plabel] at (0, 0.65) {(b) identity};
\end{scope}
\begin{scope}[xshift=6.0cm, yshift=0cm]
  \draw[pwire] (-0.8, 0.25) -- (0.8, 0.25);
  \draw[pwire] (-0.8,-0.25) -- (0.8,-0.25);
  \node[gate] at (0, 0.25) {\(K\)};
  \node[gate] at (0,-0.25) {\(L\)};
  \node[plabel] at (0, 0.65) {(c) parallel\\\(K\otimes L\)};
\end{scope}
\begin{scope}[xshift=0cm, yshift=-2.1cm]
  \draw[pwire] (-1.0, 0) -- (1.0, 0);
  \node[gate, minimum width=0.5cm] at (-0.45, 0) {\(K\)};
  \node[gate, minimum width=0.5cm] at ( 0.45, 0) {\(L\)};
  \node[plabel] at (0, 0.65) {(d) sequential\\\(L\circ K\)};
\end{scope}
\begin{scope}[xshift=3.0cm, yshift=-2.1cm]
  \draw[pwire] (-0.8, 0.2) -- (0.8, 0.2);
  \draw[pwire] (-0.4,-0.2) -- (0.8,-0.2);
  \node[prep, rotate=0] at (-0.4,-0.2) {};
  \node[font=\scriptsize, anchor=east, text=black!75] at (-0.45, -0.2) {\(\pket{0}\)};
  \node[gate, minimum height=0.7cm] at (0.2, 0) {\(K\)};
  \node[plabel] at (0, 0.65) {(e) prepare ancilla};
\end{scope}
\begin{scope}[xshift=6.0cm, yshift=-2.1cm]
  \draw[pwire] (-0.8, 0.2) -- (0.8, 0.2);
  \draw[pwire] (-0.8,-0.2) -- (0.5,-0.2);
  \draw[thick, black!70] (0.5,-0.07) -- (0.5,-0.33);
  \draw[thick, black!70] (0.6,-0.13) -- (0.6,-0.27);
  \draw[thick, black!70] (0.7,-0.18) -- (0.7,-0.22);
  \node[gate, minimum height=0.7cm] at (-0.25, 0) {\(K\)};
  \node[plabel] at (0, 0.65) {(f) discard\\(marginalise)};
\end{scope}
\end{tikzpicture}
\caption{Diagrammatic notation for stochastic kernels, matching the panel labels above. \textbf{(e)} prepares an ancilla in the fixed state \(\pket{0}\); \textbf{(f)} discards (marginalizes) a wire, denoted by a ground symbol.}
\label{fig:notation-primer}
\end{figure}
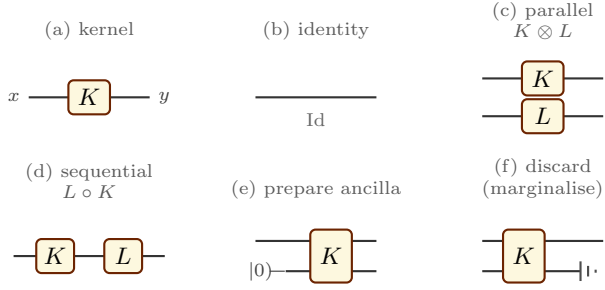

A \emph{stochastic kernel} with input space \(\mathcal{X}\) and output space
\(\mathcal{Y}\) is a function \(K:\mathcal Y\times\mathcal X\to\R\) where \(K(y\mid x)\) is the conditional
density of outputting \(y\) given input \(x\).
We write \(y\sim K(\cdot\mid x)\) for a draw from
this conditional.  
It must satisfy positivity \(K(y\mid x)\geq 0\) and
normalization for each fixed \(x\),
\begin{equation}\label{eq:kernel-norm}
  \int_{\mathcal{Y}} K(y\mid x)\,dy \;=\; 1,
\end{equation}
where the integral is a finite sum when \(\mathcal{Y}\) is discrete.
In the discrete case, \eqref{eq:kernel-norm} is equivalent to
\emph{column-stochasticity}: each column of the matrix
\(\bigl(K(y\mid x)\bigr)_{y,x}\) sums to \(1\). A stochastic kernel acts on distributions by
\begin{equation}\label{eq:kernel-action}
  \rho_{\mathrm{out}}(y)
  = (K\,\rho_{\mathrm{in}})(y)
  = \int_{\mathcal{X}} K(y\mid x)\,\rho_{\mathrm{in}}(x)\,dx,
\end{equation}
which for discrete states is the matrix vector product
\(\rho_{\mathrm{out}}(y)=\sum_{x}K(y\mid x)\,\rho_{\mathrm{in}}(x)\).

For any bounded measurable function \(f:\mathcal{X}\to\mathbb{R}\), the
expectation with respect to \(\rho\) is written uniformly as
\begin{equation}\label{eq:expectation-unified}
  \langle f,\rho\rangle
  \;=\;
  \begin{cases}
    \displaystyle\sum_{a\in\mathcal{X}} f(a)\,\rho_a \;=\; f^{\top}\rho
      & \text{(discrete wire),} \\[8pt]
    \displaystyle\int_{\mathbb{R}} f(x)\,\rho(x)\,dx
      & \text{(pmode wire).}
  \end{cases}
\end{equation}
For state spaces that combine discrete and continuous factors, the joint expectation expands to a sum-integral:
\begin{equation}\label{eq:expectation-mixed}
  \langle f,\rho\rangle
  = \sum_{a\in\mathcal{X}_{\mathrm{d}}}
    \int_{\mathcal{X}_{\mathrm{c}}}
    f(a,x)\,\rho(a,x)\,dx,
\end{equation}
where \(\mathcal{X}_{\mathrm{d}}\) and \(\mathcal{X}_{\mathrm{c}}\) are the discrete and continuous parts of the state space, respectively. Applying \(K\) to \(\rho\) transforms expectations as \(\langle f,\rho\rangle\mapsto\langle f,K\rho\rangle =\langle K^{\top}f,\rho\rangle\), so a kernel acts forward on states and backward on observables. This dual action is what allows the circuit objectives of Section~\ref{subsec:diff} to be differentiated one gate at a time.

We build stochastic programs by combining single kernels as our building blocks. Figure~\ref{fig:notation-primer} fixes the diagrammatic notation used throughout. A kernel is a box on a wire, read from input \(x\) on the left to output \(y\) on the right (Figure~\ref{fig:notation-primer}(a)), with the identity kernel \(\Id(x'\mid x)=\delta_{x',x}\), which leaves the state unchanged, drawn simply as an empty wire (Figure~\ref{fig:notation-primer}(b)). We can then compose kernels either in series and in parallel.

Parallel composition (Figure~\ref{fig:notation-primer}(c)) places kernels on disjoint wires. For \(K:\mathcal{Y}\times\mathcal{X}\to\R\) and \(L:\mathcal{Y}'\times\mathcal{X}'\to\R\) acting on disjoint state spaces,
\begin{equation}\label{eq:parallel}
  (K\otimes L)(y,y'\mid x,x')
  = K(y\mid x)\;L(y'\mid x').
\end{equation}
Since each factor integrates to one, so does their product, and parallel
composition sends probability distributions to probability distributions.

Sequential composition (Figure~\ref{fig:notation-primer}(d)) places kernels one after the other. For \(K:\mathcal{Z}\times\mathcal{X}\to\R\) followed by \(L:\mathcal{Y}\times\mathcal{Z}\to\R\),
\begin{equation}\label{eq:sequential}
  (L\circ K)(y\mid x)
  = \int_{\mathcal{Z}} L(y\mid z)\,K(z\mid x)\,dz,
\end{equation}
which in the discrete case is ordinary matrix multiplication. Since each kernel integrates to one over its output, composition sends probability distributions to probability distributions, with \(\rho_{\mathrm{out}}=(L\circ K)\,\rho_{\mathrm{in}}=L(K\,\rho_{\mathrm{in}})\). In particular, the kernels with the same input and output space \(\mathcal X\) form a semigroup under composition with the identity \(I(x'\mid x)=\delta_{x'=x}\).

Finally, two operations can change the numbers of wires. Figure~\ref{fig:notation-primer}(e) prepares an \emph{ancilla}, a fresh wire initialized to a fixed state
\(\pket{0}\) and fed into a gate, and Figure~\ref{fig:notation-primer}(f) \emph{discards} a wire by marginalizing over its values. These four operations, \(\circ\), \(\otimes\), preparation, and discarding, suffice to build every
kernel considered in this paper.

\subsection{Discrete and continuous-time Markov chains}
\label{sec:markov}

By iterating the compositions introduced in Section~\ref{sec:kernels} one can generate dynamical behavior. Here we briefly review the resulting discrete-time Markov chains and their continuous-time limit, for a more complete treatment see \cite{norris97}.

The repeated application of a fixed kernel defines a \emph{discrete-time Markov chain}. Starting from \(\rho_0\), the state after \(n\) steps is
\begin{equation}
  \rho_n = K^n\,\rho_0, \qquad K^n = \underbrace{K\circ\cdots\circ K}_{n}.
\end{equation}
At the level of sample paths, \((X_n)_{n\geq 0}\) is a stochastic process whose one-step transition satisfies \(P(X_{n+1}=x_{n+1}\mid X_0=x_0,\ldots,X_n=x_n)=K(x_{n+1}\mid x_n)\), so the future depends on the past only through the current state. A distribution \(\pi\) is \emph{stationary} for \(K\) if \(K\pi=\pi\).  Under standard ergodicity conditions (irreducibility and aperiodicity in the discrete-state case), \(\rho_n\to\pi\) as \(n\to\infty\) regardless of \(\rho_0\). We call \(K\) the transition kernel of this Markov chain.  We note that in standard literature the transition kernel \(P=K^\top\) is right multiplied to the probability vector  giving \(\rho P\) unlike here where we left multiply.

Continuous-time dynamics arise from the same construction when the
step becomes infinitesimal. Consider a kernel that deviates from the identity by a small step \(\tau\),
\begin{equation}\label{eq:euler-kernel}
  K_\tau = \Id + \tau\,Q,
\end{equation}
where column-stochasticity of \(K_\tau\) requires \(Q\) to have off-diagonal entries \(Q_{ij}\geq 0\), the transition rate from state \(j\) to state \(i\), and diagonal entries \(Q_{ii}=-\sum_{j\neq i}Q_{ij}\), so that each column sums to zero. We call \(Q\) a \emph{rate matrix} (or \emph{generator}). Composing \(N\) such steps over a time \(t\) and letting \(N\to\infty\) gives
\begin{equation}\label{eq:forward-kolmogorov}
  \rho(t)
  = \lim_{N\to\infty}\bigl(\Id + Q\,t/N\bigr)^{N}\rho_0
  = e^{Qt}\,\rho_0,
\end{equation}
which is the solution of the \emph{forward Kolmogorov equation} \(\dd\rho/\dd t = Q\,\rho\). A stationary distribution satisfies \(Q\pi=0\). The family \(K_t=e^{Qt}\) is column-stochastic for every \(t\geq 0\) and forms a one-parameter semigroup within the semigroup of kernels of Section~\ref{sec:kernels}. Fixing a step size \(t\) recovers the discrete-time picture. Both the Euler kernel \(\Id+\tau Q\) and the exact exponential \(e^{Qt}\) reappear as gates. The former in the product formulas of Section~\ref{sec:trotterization}, the latter in generator-derived gates such as \(\gname{PIsing}\) in Section~\ref{sec:gates}.

To express larger parametrized probabilistic programs, we next allow many kernels to be organized in an acyclic graph and composed through their shared variables.

\subsection{Directed Factor Graphs}
\label{sec:dfg}

\begin{figure}[t]
\centering
\begin{tikzpicture}[
  >=stealth,
  var/.style={circle, draw=extBrown, thick, minimum size=8mm, inner sep=0pt,
              text=extBrown},
  fac/.style={rectangle, draw=extBrown, thick, rounded corners=2pt,
              minimum width=16mm, minimum height=7mm, inner sep=3pt,
              font=\footnotesize, text=white},
  edge/.style={->, thick, extBrown}
]
  \def\cx{0}   
  \def\ck{2}   
  \def\cz{4}   
  \def\cl{6}   
  \def\cy{8}   
  \def\rtop{1.2}
  \def\rbot{-1.2}
  \node[var, fill=extYellow] (x)  at (\cx, 0)      {\(x\)};
  \node[var, fill=extGold]   (z1) at (\cz, \rtop)  {\(z_1\)};
  \node[var, fill=extGold]   (z2) at (\cz, \rbot)  {\(z_2\)};
  \node[var, fill=extSolar]  (y1) at (\cy, \rtop)  {\(y_1\)};
  \node[var, fill=extSolar]  (y2) at (\cy, \rbot)  {\(y_2\)};
  \node[fac, fill=extCopper]  (k1) at (\ck, \rtop) {\(K_1(\theta_1)\)};
  \node[fac, fill=extCopper]  (k2) at (\ck, \rbot) {\(K_2(\theta_2)\)};
  \node[fac, fill=extFuchsia] (k3) at (\cl, \rtop) {\(K_3(\theta_3)\)};
  \node[fac, fill=extFuchsia] (k4) at (\cl, \rbot) {\(K_4(\theta_4)\)};
  \draw[edge] (x)  -- (k1);
  \draw[edge] (k1) -- (z1);
  \draw[edge] (x)  -- (k2);
  \draw[edge] (k2) -- (z2);
  \draw[edge] (z1) -- (k3);
  \draw[edge] (k3) -- (y1);
  \draw[edge] (z2) -- (k4);
  \draw[edge] (k4) -- (y2);
\end{tikzpicture}
\caption{A directed factor graph of stochastic kernels. Variable nodes (circles) are connected via factor nodes (rectangles) representing the kernels \(K_i(\theta_i)\).}
\label{fig:param-kernel-dag}
\end{figure}
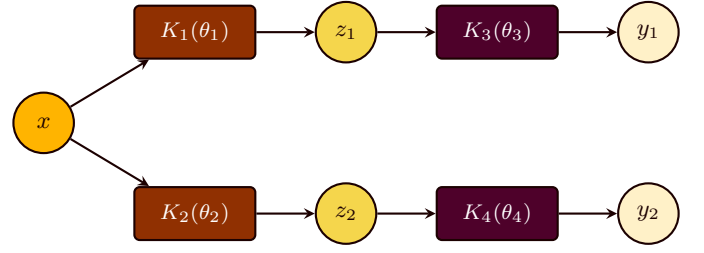

The compositions of Section~\ref{sec:kernels} build kernels in series and in parallel, but many probabilistic models are specified more loosely, as a network of conditional distributions in which several kernels consume and produce shared variables. A Directed Factor Graph (DFG) captures this general case: it is a directed acyclic graph of stochastic kernels, where the output of each kernel is fed as input to one or more subsequent kernels. This structure is an instance of a directed graphical model~\cite{bishop2006pattern}, where nodes represent stochastic kernels, directed edges carry the variables passed between them, and internal variables are marginalized out to yield an overall conditional distribution from the designated inputs to the designated outputs. We use the DFG as the internal representation underlying the PSCs of Section~\ref{sec:stoc_circs}, and to derive the gradient estimators of Section~\ref{subsec:diff}.

Each kernel \(K_i(\theta_i) : \mathcal{Y} \times \mathcal{X} \to \mathbb{R}\) is indexed by parameters \(\theta_i \in \Theta_i \subseteq \mathbb{R}^{d_i}\), with entries \(K_i(y \mid x; \theta_i)\). We shall sometimes omit the dependence on \(\theta_i\) when convenient. We call \(K_i\) a \emph{parent} of \(K_j\) if there is a directed edge from \(K_i\) to \(K_j\), i.e.\ an output of \(K_i\) serves as an input to \(K_j\). We assume a hierarchical (also called topological) ordering: all parents appear before their children, so if \(K_i\) is a parent of \(K_j\), then \(i<j\).
This ordering means the DAG can always be evaluated as the
sequential composition
\begin{equation}\label{eq:dfg-sequential}
  \mathcal{K}(\theta) = K_n(\theta_n) \circ \cdots \circ K_1(\theta_1),
\end{equation}
parametrized by \(\theta = (\theta_1, \ldots, \theta_n) \in \Theta\) and the state evolves through the graph as
\begin{align}
  \rho_j           &= K_j\,\rho_{j-1}, \notag \\
  \rho_{\mathrm{out}} &= \mathcal{K}\,\rho_{\mathrm{in}}
                      = K_n \cdots K_1\,\rho_{\mathrm{in}}.
  \label{eq:dfg-evolution}
\end{align}
The corresponding sample path is \(z^{(0)},\ldots,z^{(n)}\) with
\begin{align*}
  z^{(0)} &\sim \rho_{\mathrm{in}}, \\
  z^{(j)} &\sim K_{j}(\,\cdot \mid z^{(j-1)}), \qquad j=1,\ldots,n,
\end{align*}
so that \(z^{(n)}\sim\rho_{\mathrm{out}}\). We call \(z^{(0)}\) the input, \(z^{(n)}\) the output, and \(z^{(1)},\ldots,z^{(n-1)}\) the latent variables. We note that, as shown in Figure~\ref{fig:param-kernel-dag}, the same DAG may admit several hierarchical orderings. We do not assume access to a closed-form expression for the conditional distribution, we assume only that each kernel \(K_j\) can be sampled efficiently. A smooth parameterization is typically chosen in learning settings so that the gradients \(\nabla_{\theta_i} K_i(y \mid x)\) can be computed.

A DFG places no constraint on how kernels are scheduled beyond the topological ordering. The PSC of Section~\ref{sec:stoc_circs} equips this structure with a fixed register of typed wires and ordered layers of gates with disjoint supports. The layered structure makes parallel execution explicit, thus making compilation to hardware more obvious. Conversely, every PSC unrolls to a DFG by treating each gate as a kernel and each wire segment as a shared variable.

\section{Parametrized Stochastic Circuits}\label{sec:stoc_circs}

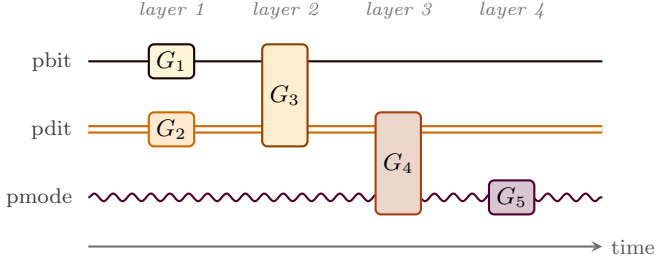
\begin{figure}[t]
\centering
\begin{tikzpicture}[
  >=stealth,
  font=\small,
  pbitwire/.style={thick, extBrown},
  pditwire/.style={thick, extOrange!80!black, double, double distance=1.6pt},
  pmodewire/.style={thick, extFuchsia, decorate, decoration={snake, amplitude=0.5mm, segment length=2.5mm}},
  pbitgate/.style={draw, thick, fill=extGold!22, draw=extBrown,
                   rounded corners=2pt, inner sep=1pt},
  pditgate/.style={draw, thick, fill=extOrange!18, draw=extOrange!80!black,
                   rounded corners=2pt, inner sep=1pt},
  pmodegate/.style={draw, thick, fill=extFuchsia!22, draw=extFuchsia,
                    rounded corners=2pt, inner sep=1pt},
  pbitpditgate/.style={draw, thick, fill=extGold!50!extOrange!22, draw=extBrown!50!extOrange,
                       rounded corners=2pt, inner sep=1pt},
  pditpmodegate/.style={draw, thick, fill=extOrange!50!extFuchsia!22, draw=extOrange!50!extFuchsia,
                        rounded corners=2pt, inner sep=1pt},
  wlabel/.style={font=\footnotesize, anchor=east, text=black!75},
  layerlabel/.style={font=\scriptsize\itshape, text=black!55, anchor=south},
]
\draw[pbitwire]  (-0.4, 1.8) -- (6.4, 1.8);
\draw[pditwire]  (-0.4, 0.9) -- (6.4, 0.9);
\draw[pmodewire] (-0.4, 0.0) -- (6.4, 0.0);

\node[wlabel] at (-0.5, 1.8) {pbit};
\node[wlabel] at (-0.5, 0.9) {pdit};
\node[wlabel] at (-0.5, 0.0) {pmode};

\node[pbitgate,  minimum width=0.6cm, minimum height=0.45cm] at (0.7, 1.8) {\(G_1\)};
\node[pditgate,  minimum width=0.6cm, minimum height=0.45cm] at (0.7, 0.9) {\(G_2\)};

\node[pbitpditgate, minimum width=0.6cm, minimum height=1.35cm] at (2.2, 1.35) {\(G_3\)};

\node[pditpmodegate, minimum width=0.6cm, minimum height=1.35cm] at (3.7, 0.45) {\(G_4\)};

\node[pmodegate, minimum width=0.6cm, minimum height=0.45cm] at (5.2, 0.0) {\(G_5\)};

\node[layerlabel] at (0.7,  2.25) {layer 1};
\node[layerlabel] at (2.2,  2.25) {layer 2};
\node[layerlabel] at (3.7,  2.25) {layer 3};
\node[layerlabel] at (5.2,  2.25) {layer 4};

\draw[->, thick, black!55] (-0.4, -0.65) -- (6.4, -0.65);
\node[font=\footnotesize, text=black!65, anchor=west] at (6.4, -0.65) {time};
\end{tikzpicture}
\caption{Within layer 1, gates \(G_1\) and \(G_2\) execute in parallel. Layer 2 contains a cross-type gate \(G_3\) coupling the pbit and the pdit. Layer 3 contains a cross-type gate \(G_4\) coupling the pdit and the pmode. Layer 4 acts on the pmode.}
\label{fig:psc-example}
\end{figure}

A Parametrized Stochastic Circuit is the circuit-shaped specialization of the directed factor graph in Section~\ref{sec:dfg}. Its kernels are grouped into ordered layers: kernels in the same layer have disjoint state spaces and execute in parallel, while layers compose in series to define a global stochastic map on the full state space. Within a PSC, wires carry states of the data types described in Sec.~\ref{sec:notation}. All stochasticity is localized to the gate kernels, while the wires themselves only transmit the variables' state between layers. PSCs give a hardware-near model for stochastic computation. A logical circuit defined within this model can be compiled to backends with different execution rules, such as full state-space simulation and sample-based execution with software-sourced pseudo-randomness and hardware-sourced randomness (via the XTR-0 platform, Sec.~\ref{sec:hard-disc}). The companion manuscript~\cite{amico2026thermalizing} also explores execution of PSCs on thermodynamic hardware (the Z1 chip of Fig.~\ref{fig:z1-arch}) by compiling them to an energy-based model intermediate representation.

We now give more precise definitions of the components of a PSC. A \emph{register} is the full collection of wires acted on by the circuit. Using the notation of Section~\ref{sec:notation}, let \(\mathcal W\) be the set of wires in the register, \(\mathcal X_w\) the state space of wire \(w\), and \(\mathcal X_I=\prod_{w\in I}\mathcal X_w\) the product state space of a subset \(I\subseteq\mathcal W\). A local stochastic gate supported on \(I\) is a stochastic kernel
\begin{equation}
    G_I(\theta):\mathcal X_I\times\mathcal X_I\to\R,
\end{equation}
with tunable parameters \(\theta\). When an operation changes the number of wires, it is represented in the fixed-register picture by preparing ancilla wires and discarding unused wires, using the primitives of Section~\ref{sec:kernels}. We define a set of \emph{elementary} gates in Section~\ref{sec:gates}. These are the primitives from which more complex kernels are built by composition in series and parallel. 

A \emph{layer} is a collection of gates with pairwise disjoint supports, as illustrated in Figure~\ref{fig:psc-example}, where gates \(G_1\) and \(G_2\) occupy the same layer and execute in parallel. If \(J=\{I_1,\ldots,I_m\}\) is the collection of active supports in the layer, then
\begin{equation}
    L_J(\theta) = \bigotimes_{r=1}^m G_{I_r}(\theta_r).
\end{equation}
Wires outside the active support pass unchanged to the next layer:
\begin{equation}
    \widetilde L_J(\theta)=L_J(\theta)\otimes \mathbb I_{(\bigcup_{r=1}^m I_r)^c}.
\end{equation}
This embedding, together with any necessary wire permutation, is understood implicitly. A \emph{parametrized stochastic circuit}
\begin{equation}
    \mathcal G(\theta)
    =
    \widetilde L_{J_k}(\theta_k)\cdots
    \widetilde L_{J_1}(\theta_1):\mathcal X_{\mathcal W}\times \mathcal X_{\mathcal W}\to\R,
    \label{eq:psc-operator-program}
\end{equation}
is a stochastic kernel on the register \(\mathcal X_{\mathcal W}\): the register's state is a distribution \(\rho\) over \(\mathcal X_{\mathcal W}\), while any single realization (sample) of the register takes a definite, deterministic value in \(\mathcal X_{\mathcal W}\).
It is composed of layers \(\widetilde L_{J_j}(\theta_j)\), and the composition
is read from right to left: \(\widetilde L_{J_1}(\theta_1)\) acts first, then
\(\widetilde L_{J_2}(\theta_2)\), and so on until
\(\widetilde L_{J_k}(\theta_k)\) acts last.

\subsection{Elementary Gates}\label{sec:gates}

Gates of a PSC are defined here through three possible constructions. The first is the direct specification of a stochastic matrix. The second is via convex combinations of deterministic operations. For any deterministic gate \(B\), the mixture \((1-p)\,\Id + p\,B\) is implemented by preparing an auxiliary pbit in the state \(\rho_{\mathrm{aux}}=(1-p)\pket{0}+p\pket{1}\), applying \(B\) conditioned on the auxiliary state \(\pket{1}\) and nothing conditioned on \(\pket{0}\), and discarding the auxiliary pbit. A convenient parametrization for learning passes the mixture weight through a sigmoid,
\begin{equation}\label{eq:sigmoid-gate}
    G(\theta) = \bigl(1-\sigma(\theta)\bigr)\,\Id + \sigma(\theta)\,B,
    \qquad
    \sigma(\theta)=\frac{1}{1+e^{-\theta}},
\end{equation}
so that \(p=\sigma(\theta)\in(0,1)\) for every \(\theta\in\R\). This is the gate family to which the parameter-shift rule of Section~\ref{subsec:diff} applies. The third construction method uses exponentials of the local generators of Section~\ref{sec:markov}. Continuous-time finite-state Markov models are generated by rate matrices whose exponentials are stochastic transition maps~\cite{norris97}. To define a gate, choose a finite-state rate matrix \(Q\) and exponentiate it to obtain a generator-derived stochastic kernel \(K(t)=\exp(tQ)\). Markov-type Lie theory provides a reference point for such generator-based operator families~\cite{johnson1985markov}. In the generator-derived subfamily, PSCs recover the operator-splitting picture where local Markov generators are exponentiated to finite-time stochastic kernels and then composed in time. 

Let us now define some elementary gates for stochastic circuits, organized by wire type (further gates are cataloged in the Appendices~\ref{app:gates} and~\ref{app:pmode}).

\paragraph{pbit and pdit}
The probabilistic NOT (\(\gname{PNOT}\)) flips the state of a pbit with probability \(p \in [0,1]\):
\begin{equation}
    \gname{PNOT}(p) = \begin{pmatrix}
        1 - p & p \\
        p & 1 - p
    \end{pmatrix}.
\end{equation}
At \(p=0\) this is the identity, at \(p = 1\) it is the deterministic \(\gname{NOT}\), and at \(p = \tfrac{1}{2}\) it produces a uniform distribution regardless of input. Applying \(\gname{PNOT}(p)\) to a single-pbit with distribution \(\rho = (\alpha,\, \beta)^\top\) yields
\begin{equation}
    \gname{PNOT}(p)\,\rho
    = \begin{pmatrix}
        (1-p)\,\alpha + p\,\beta \\
        p\,\alpha + (1-p)\,\beta
    \end{pmatrix},
\end{equation}
exchanging probability mass between \(\pket{0}\) and \(\pket{1}\).

Gates can also act on multiple binary wires, the probabilistic swap (\(\gname{PSWAP}\)) is one such example.  It operates on two pbits, exchanging their states with probability \(p \in [0,1]\):
\begin{equation}
    \gname{PSWAP}(p)\,\pket{ab} = (1-p)\,\pket{ab} + p\,\pket{ba}.
\end{equation}
Like \(\gname{PNOT}\), \(\gname{PSWAP}\) is a convex combination of the identity and a deterministic operation.

As an example of a generator-derived gate we take \(\gname{PIsing}\), which defines a local two-spin Glauber kernel~\cite{levin2007glauberdynamicsmeanfieldising}. Given parameters \(\theta = (J, h_1, h_2, \beta, \Delta t)\), the energy of a two-pbit configuration \(\mathbf{s} = (s_1, s_2) \in \{0,1\}^2\) is
\begin{equation}
    E(\mathbf{s}) = -J(2s_1-1)(2s_2-1) - h_1(2s_1-1) - h_2(2s_2-1).
\end{equation}
The associated Glauber dynamics has rate matrix \(Q\) with off-diagonal entries
\begin{equation}
    Q_{ab} = \sigma\!\left[-\beta\bigl(E(\mathbf{s}_a)-E(\mathbf{s}_b)\bigr)\right]
\end{equation}
for single-spin-flip transitions \(a \neq b\), zero otherwise, and diagonal \(Q_{bb} = -\sum_{a\neq b} Q_{ab}\). The gate is
\begin{equation}
    \gname{PIsing}(\theta) = \exp(\Delta t\,Q),
\end{equation}
which is column-stochastic for \(\Delta t \geq 0\).
We expand upon this in our companion paper~\cite{amico2026thermalizing} which focuses exclusively on energy-based transition kernels.

\paragraph{pmode}
A pmode gate is a kernel \(K(y\mid x)\) on \(\R^N\). The elementary directly specified gate is the \emph{affine Gaussian gate} \((M,\bm d,\Delta)\),
\begin{equation}
X \mapsto MX + \bm{d} + \varepsilon, \qquad X \sim \mathcal{N}(\bm{\mu}, \Sigma),\quad \varepsilon \sim \mathcal{N}(0, \Delta),
\label{eq:pmode-gaussian-update}
\end{equation}
which maps \(\mathcal N(\bm\mu,\Sigma)\) to \(\mathcal N(M\bm\mu+\bm d,\,M\Sigma M^{\top}+\Delta)\), and special cases such as displacements, scalings, and rotations are cataloged in Appendix~\ref{app:pmode}. The generator-derived example is the Ornstein--Uhlenbeck gate \(\gname{POU}(\gamma,D,t)\), obtained by exponentiating the Fokker--Planck generator of the SDE \(\dd X_t=-\gamma X_t\,\dd t+\sqrt{2D}\,\dd W_t\); it combines deterministic contraction with noise injection and plays for pmodes the role \(\gname{PIsing}\) plays for pbits (Appendix~\ref{app:pmode}). These Gaussian primitives support a broad family of continuous-variable algorithms, ranging from Gaussian sampling to deep neural networks, whose execution on equilibrium-based superconducting thermodynamic hardware is demonstrated in Ref.~\cite{lockwood2026blueprint}, and could be readily implemented in CMOS using X0 or in Z1 via \texttt{thermalizers} compilation~\cite{amico2026thermalizing}.

\paragraph{Hybrid pdit/pmode gates}
Many stochastic systems couple a discrete state to a continuous one. PSCs accommodate this through gates acting on the product space \(\{0,\ldots,d-1\}\times\R^{N}\) with the compositional rules of Section~\ref{sec:kernels} unchanged.

Controlled shift \(\gname{CShift}(\bm\alpha)\): a pdit conditions a deterministic displacement on the pmode,
\begin{equation}\label{eq:cshift-gate}
    \gname{CShift}(\bm\alpha):\;\pket k\otimes p(x)\;\longmapsto\;\pket k\otimes p(x-\alpha_k),
\end{equation}
where \(\bm\alpha=(\alpha_0,\ldots,\alpha_{d-1})\) gives the shift selected by each pdit state. 
The pmode-side action when the pdit is held in state \(k\) is a \(\gname{PDisp}(\alpha_k)\), the displacement gate defined in Appendix~\ref{app:pmode}.

Controlled scale \(\gname{CScale}(\bm r)\): the multiplicative analog,
\begin{equation}\label{eq:cscale-gate}
    \gname{CScale}(\bm r):\;\pket k\otimes p(x)\;\longmapsto\;\pket k\otimes e^{-r_k}p(e^{-r_k}x),
\end{equation}
rescales the pmode by \(e^{r_k}\) when the pdit is in state \(k\).

Mixture-of-Gaussians \(\gname{MoG}(\bm\pi,\bm\alpha)\): preparing a pdit ancilla with probability distribution \(\bm\pi=(\pi_0,\ldots,\pi_{d-1})\in\Delta_{d-1}\), applying \(\gname{CShift}(\bm\alpha)\) to a Gaussian pmode \(X\sim\mathcal N(\mu,\sigma^2)\), and marginalizing the ancilla yields the Gaussian mixture
\begin{equation}\label{eq:mog-mixture}
    p_{X'}(x) = \sum_{k=0}^{d-1}\pi_k\,\mathcal N(x;\mu+\alpha_k,\sigma^2),
\end{equation}
a universal density approximator under standard regularity conditions. Combining \(\gname{CScale}\) with \(\gname{CShift}\) yields the heteroscedastic variant \(\sum_k\pi_k\,\mathcal N(e^{r_k}\mu+\alpha_k,\,e^{2r_k}\sigma^2)\). Note that \(\gname{MoG}\) is precisely the construction of Figure~\ref{fig:notation-primer}(e)--(f): a prepared ancilla, a controlled gate, and a discard.

\subsection{Trotterization}\label{sec:trotterization}

In Section~\ref{sec:gates} we have introduced gates that act locally on a few wires at a time. However, in most cases one is interested in the dynamics generated by a sum of local terms
\begin{equation}\label{eq:generator-sum-rewrite}
    Q = \sum_{j=1}^{k} Q_j,
\end{equation}
where each \(Q_j\) is supported on a few wires while the exponential \(e^{Qt}\) is a kernel on the full register, in general not equal to any product of local gates.

\emph{Trotterization}~\cite{trotter1959product} allows us to approximate \(e^{Qt}\) by a circuit of local gates applied over \(N\) steps of size \(\tau=t/N\), each implementing either the exact local exponential \(e^{Q_j\tau}\) or the Euler kernel \(\Id+\tau Q_j\) of Eq.~\eqref{eq:euler-kernel}. For example, in Section~\ref{sec:graph-diff} the graph Laplacian decomposes into one generator per edge, and each factor is a single \(\gname{PSWAP}\) gate.

Given access to exact local exponentials, one can approximate \(e^{Qt}\) via \emph{Lie--Trotter splitting}~\cite{trotter1959product}, which splits the exponential of the sum into a product of local exponentials,
\begin{equation}\label{eq:lie-trotter}
    e^{Qt} = \left(e^{Qt/N}\right)^{\!N}
    \approx \left(\prod_{j=1}^k e^{Q_jt/N}\right)^{\!N}.
\end{equation}
The error of a single step is controlled by the commutators \([Q_i,Q_j]\), and vanishes when all generators commute. The single-step error is of order \(O(t^2k^2/N^2)\), and since the errors of the \(N\) steps accumulate at most additively (the telescoping bound), the global error is \(O(t^2k^2/N)\). This is a first-order method, doubling \(N\) halves the global error. The splitting error can be reduced by symmetrizing the product. The second-order \emph{Strang splitting}~\cite{strang1968construction} approximates
\begin{align}
    e^{Qt/N} \approx &e^{Q_1t/2N}e^{Q_2t/2N}\cdots e^{Q_{k-1}t/2N}e^{Q_kt/N}\label{eq:strang}
    \\&e^{Q_{k-1}t/2N}\cdots e^{Q_2t/2N}e^{Q_1t/2N},\notag
\end{align}
which cancels the leading commutator error by symmetry and achieves a local error of \(O(t^3k^3/N^3)\), hence a global error of \(O(t^3k^3/N^2)\), at the cost of \(2k-1\) rather than \(k\) gate applications per step. Higher-order product formulas~\cite{suzuki1990fractal} exist but are not needed here.

When the exact local exponentials are not available as gates, each factor can be replaced by its Euler kernel,
\begin{equation}\label{eq:trotter-product-formula}
    e^{Qt} = \left(e^{Qt/N}\right)^{\!N}
    \approx \left(\prod_{j=1}^{k} \bigl(\Id + Q_j\,t/N\bigr)\right)^{\!N},
\end{equation}
and the single-step error decomposes into an Euler part and a splitting part,
\begin{align*}
    &\left\lVert e^{Qt/N} - \prod_{j=1}^k\!\bigl(\Id + Q_j\,t/N\bigr)\right\rVert
    \leq
    \underbrace{\left\lVert e^{Qt/N} - \bigl(\Id + Qt/N\bigr)\right\rVert}_{O(t^2/N^2)}
    \\+&
    \underbrace{\left\lVert \bigl(\Id + Qt/N\bigr) - \prod_{j=1}^k\!\bigl(\Id + Q_j\,t/N\bigr)\right\rVert}_{O(t^2 k^2/N^2)},
\end{align*}
giving the same \(O(t^2k^2/N)\) global error as Lie--Trotter. These bounds hold for any submultiplicative matrix norm; the implicit constant is proportional to the maximum pairwise commutator norm \(\max_{i\neq j}\|[Q_i,Q_j]\|\).

The number of factors \(k\) entering these bounds can often be reduced by grouping. If some of the generators commute pairwise, one constructs the non-commutation graph on \(\{Q_j\}\) and colors it with \(\chi\) colors, so that generators of the same color commute. Within a color class the product is exact, \(e^{G_c\tau}=\prod_{j\in\mathcal C_c}e^{Q_j\tau}\) with \(G_c=\sum_{j\in\mathcal C_c}Q_j\), so Trotterizing the \(\chi\) grouped generators replaces \(k\) by \(\chi\) in all the bounds above. For generators tiled over the edges of a graph this is an edge coloring, the case used in Section~\ref{sec:graph-diff}.

In practice the choice of formula reduces to what is available as a gate. When only Euler kernels \(\Id+Q_j\tau\) are available, both splittings incur an additional \(O(kt^2/N)\) Euler error that dominates at any practical \(N\), so Lie--Trotter is simpler and sufficient. When exact local exponentials are available, Strang is preferable for \(N\gtrsim kt\) steps, where its error \(O(k^3t^3/N^2)\) drops below the Lie--Trotter error \(O(k^2t^2/N)\).

\subsection{Differentiation \& Optimization}\label{subsec:diff}

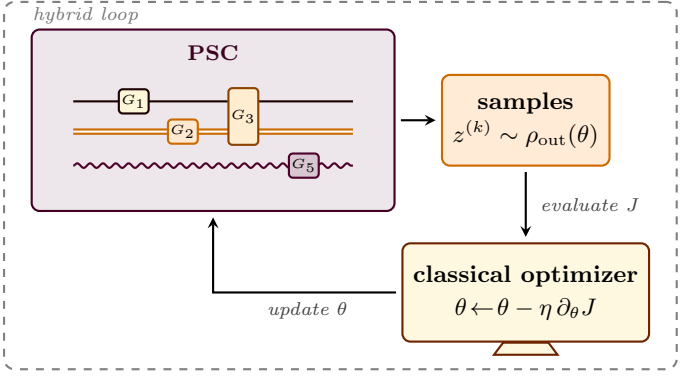
\begin{figure}[t]
\centering
\begin{tikzpicture}[
  >=stealth,
  font=\small,
  panelP/.style={draw=extFuchsia!85!black, thick, rounded corners=3pt,
                 fill=extFuchsia!10, inner sep=5pt},
  panelR/.style={draw=extOrange!80!black, thick, rounded corners=3pt,
                 fill=extOrange!16, inner sep=4pt, align=center,
                 minimum width=2.0cm, minimum height=1.2cm},
  panelC/.style={draw=extCopper!75!black, thick, rounded corners=3pt,
                 fill=extGold!18, inner sep=4pt, align=center,
                 minimum width=2.5cm, minimum height=1.3cm},
  minigate/.style={draw, thick, rounded corners=1.5pt, inner sep=1pt,
                    minimum width=0.32cm, minimum height=0.32cm, font=\tiny},
  bigarr/.style={->, thick, shorten >=2pt, shorten <=2pt},
  alab/.style={font=\scriptsize\itshape, text=black!75},
  hybridlbl/.style={font=\scriptsize\itshape, text=black!55,
                    fill=white, inner sep=1pt},
]
\node[panelP, minimum width=4.8cm, minimum height=2.4cm] (psc) at (0, 0) {};
\node[font=\footnotesize\bfseries, anchor=north, text=extFuchsia!75!black]
     at ([yshift=-3pt]psc.north) {PSC};
\begin{scope}[shift={(psc.center)}, yshift=-0.20cm]
  \draw[thick, extBrown] (-1.85, 0.45) -- (1.85, 0.45);
  \draw[thick, extOrange!80!black, double, double distance=0.8pt]
        (-1.85, 0.05) -- (1.85, 0.05);
  \draw[thick, extFuchsia, decorate,
        decoration={snake, amplitude=0.3mm, segment length=1.7mm}]
        (-1.85,-0.40) -- (1.85,-0.40);
  \node[minigate, fill=extGold!22, draw=extBrown] at (-1.05, 0.45) {\(G_1\)};
  \node[minigate, fill=extOrange!18, draw=extOrange!80!black] at (-0.4, 0.05) {\(G_2\)};
  \node[minigate, fill=extGold!50!extOrange!22, draw=extBrown!50!extOrange,
        minimum height=0.75cm] at (0.4, 0.25) {\(G_3\)};
  \node[minigate, fill=extFuchsia!22, draw=extFuchsia] at (1.2,-0.40) {\(G_5\)};
\end{scope}

\node[panelR, right=0.6cm of psc] (ro)
     {\textbf{samples}\\[2pt]\(z^{(k)}\sim\rho_{\mathrm{out}}(\theta)\)};

\node[panelC, below=1.0cm of ro] (opt)
     {\textbf{classical optimizer}\\[2pt]
      \(\theta\!\gets\!\theta-\eta\,\partial_\theta J\)};
\draw[draw=extCopper!75!black, thick, fill=extGold!18]
     ($(opt.south)+(-0.30,0)$) -- ($(opt.south)+(0.30,0)$)
     -- ($(opt.south)+(0.42,-0.16)$) -- ($(opt.south)+(-0.42,-0.16)$) -- cycle;

\draw[bigarr] (psc.east) -- node[above, alab] {} (ro.west);
\draw[bigarr] (ro.south) -- node[right, alab, xshift=2pt] {evaluate \(J\)} (opt.north);
\draw[bigarr] (opt.west) -| node[below, alab, pos=0.25] {update \(\theta\)}
              (psc.south);

\node[draw=black!50, thick, dashed, rounded corners=4pt,
      fit=(psc)(ro)(opt), inner sep=10pt] (hybridbox) {};
\node[hybridlbl, anchor=north west]
     at ([xshift=10pt, yshift=-1pt]hybridbox.north west) {hybrid loop};
\end{tikzpicture}
\caption{The PSC (top-left) is sampled to produce samples \(z^{(k)}\sim\rho_{\mathrm{out}}(\theta)\) (top-right); a classical optimizer (bottom-right) evaluates the scalar loss \(J(\theta)\) from those samples and updates \(\theta\) for the next iteration.}
\label{fig:hybrid-loop}
\end{figure}

Here we focus on methods for training a PSC by gradient descent on a scalar objective estimated from samples, which requires gradient estimators that are themselves computable by running circuits. Figure~\ref{fig:hybrid-loop} illustrates this hybrid classical-stochastic loop. We present three such estimators, in increasing order of the structure they assume on the gates: the REINFORCE estimator, valid for any smoothly parametrized gate; the parameter-shift rule, valid for the sigmoid-mixture gates of Eq.~\eqref{eq:sigmoid-gate}; and an energy-based estimator, valid for gates sampling from a Boltzmann conditional. Following Section~\ref{sec:dfg}, we derive them at the generality of the underlying DFG. We write the circuit as gates \(G_j\equiv G_{I_j}(\theta_j)\), \(j=1,\ldots,k\), applied in topological order; running the circuit once produces a latent trajectory \((z^{(0)},\ldots,z^{(k)})\) with \(z^{(0)}\sim\rho_{\mathrm{in}}\) and \(z^{(j)}\sim G_j(\,\cdot\mid z^{(j-1)})\), so that \(z^{(k)}\sim\rho_{\mathrm{out}}(\theta)\).

The objective is the expectation of a readout function \(f\) under the output distribution \(\rho_{\mathrm{out}}(\theta)=G_k\cdots G_1\,\rho_{\mathrm{in}}\),
\begin{align}
    J(\theta)
    = \mathbb E_{Y\sim\rho_{\mathrm{out}}(\theta)}[f(Y)]
    &= \langle f,\rho_{\mathrm{out}}(\theta)\rangle \nonumber \\
    &= \langle f,\; G_k\cdots G_1\,\rho_{\mathrm{in}}\rangle,
    \label{eq:readout-objective}
\end{align}
which we want to optimize with respect to \(\theta\). Since every \(G_j\) is linear in the state, the derivative follows from the product rule,
\begin{equation}
    \partial_{\theta_j}J
    =
    \langle f,\; G_k\cdots G_{j+1}\,(\partial_{\theta_j}G_j)\,
    G_{j-1}\cdots G_1\,\rho_{\mathrm{in}}\rangle,
    \label{eqn:optimize_general_exp}
\end{equation}
where the dual action of Section~\ref{sec:kernels} moves the fixed gates onto the observable side. Writing the derivative of a kernel entrywise as
\((\partial_{\theta_j}G_j)(y\mid x)
 = G_j(y\mid x)\,\partial_{\theta_j}\log G_j(y\mid x)\)
turns Eq.~\eqref{eqn:optimize_general_exp} into an expectation over trajectories,
\begin{equation}
    \partial_{\theta_j}J
    = \mathbb{E}_{(z^{(0)},\ldots,z^{(k)})}\!\left[
      f(z^{(k)})\,
      \partial_{\theta_j}\log G_j(z^{(j)}\mid z^{(j-1)})\right],
    \label{eqn:reinforce}
\end{equation}
the REINFORCE (or score-function) estimator~\cite{williams1992simple}: drawing \(S\) independent trajectories and averaging the bracket gives an unbiased estimate of the full gradient from a single batch of circuit runs. The entrywise identity requires \(G_j(z^{(j)}\mid z^{(j-1)})>0\) on the support of the trajectory distribution, which holds for the gate library of Section~\ref{sec:gates} provided parameter values are kept away from the boundary \(p\in\{0,1\}\). In the case of the sigmoid parametrization of Eq.~\eqref{eq:sigmoid-gate}, this is enforced automatically.

\paragraph{Parameter shift rule estimator}
Consider the sigmoid-mixture gates \(G(\theta)=(1-\sigma(\theta))\,\Id+\sigma(\theta)\,B\) of Eq.~\eqref{eq:sigmoid-gate}. Taking derivatives,
\begin{equation}\label{eq:grad-inf}
    \frac{\partial G(\theta)}{\partial\theta}
    = \sigma'(\theta)\,(B-\Id).
\end{equation}
Since \(\sigma'(\theta)=\sigma(\theta)\bigl(1-\sigma(\theta)\bigr)\), the derivative is itself a difference of two gates of the same family. Defining the shifted parameter \(\theta^{-}\) by \(\sigma(\theta^{-})=\sigma(\theta)^{2}\), that is
\begin{equation}\label{eq:shifted-param}
    \theta^{-}=-\ln\!\bigl((1+e^{-\theta})^{2}-1\bigr),
\end{equation}
an explicit computation gives the logarithmic shift identity
\begin{equation}\label{eq:grad-log}
    \frac{\partial G(\theta)}{\partial\theta}
    = G(\theta)-G(\theta^{-}),
\end{equation}
which expresses \(\partial_{\theta}G\) as a difference of two executable gates.

If gate \(G_j\) is of the specified form, using~\eqref{eqn:optimize_general_exp} we can express the gradient as a difference of two circuit expectations,
\begin{equation}
    \frac{\partial \langle f,\rho_{\mathrm{out}}(\theta)\rangle}{\partial\theta_j}\bigg|_{\theta_j=\theta_j^*}
    =\mathbb E_{X\sim \rho(\theta^*)}[f(X)]-\mathbb E_{\widetilde X\sim \rho(\theta^{*-})}[f(\widetilde X)],
\end{equation}
where \(\rho(\theta^*)\) is the output distributions induced by running the original circuit and \(\rho(\theta^{*-})\) the shifted circuit with parameter \(\theta_j\) changed to,
\begin{equation}
    \theta_j^{*-}=-\ln\!\bigl((1+e^{-\theta_j^*})^2-1\bigr).
\end{equation}
Drawing \(S\) iid samples \(X_i\sim \rho_{\mathrm{out}}(\theta^*)\) and \(\widetilde X_i\sim \rho_{\mathrm{out}}(\theta^{*-})\) gives the unbiased estimators
\begin{equation}
    \widehat g_j
    =
    \frac{1}{S}\sum_{i=1}^S f(X_i)
    -
    \frac{1}{S}\sum_{i=1}^S f(\widetilde X_i),
    \label{fn_grad_backprop}
\end{equation}
with variance \(\operatorname{Var}(\widehat g_j)=(\operatorname{Var}_{p_j^+}\!f+\operatorname{Var}_{p_j^-}\!f)/S\). 
The rule has two limitations. First, it applies only to gates of the sigmoid-mixture form of Eq.~\eqref{eq:sigmoid-gate}. Second, each parameter requires a separate shifted circuit evaluation, so unlike REINFORCE the full gradient cannot be obtained from a single batch of runs. However, only the gates downstream of \(G_j\) in the DFG need to be resampled; the remaining samples of the latent trajectory can be reused.

\paragraph{Energy-Based Kernels}

Suppose one of the kernels \(G(\theta)\) samples from a Boltzmann distribution, i.e., the conditional distribution \(G(\theta)(y\mid x)\) is obtained by conditioning a joint distribution with density
\begin{equation}
    p_\theta(x,y)
    =
    \frac{1}{Z(\theta)}e^{-E(x,y;\theta)}.
    \label{eq:ebm-joint}
\end{equation}
For a fixed input \(x\), the resulting kernel is
\begin{equation}
    G(\theta)(y\mid x)
    =
    \frac{e^{-E(x,y;\theta)}}{Z(x;\theta)},
    \qquad
    Z(x;\theta)
    =
    \int_{\mathcal Y} e^{-E(x,y';\theta)}\,\dd y',
    \label{eq:ebm-conditional}
\end{equation}
where the integral is replaced by a sum when \(\mathcal Y\) is discrete.

We can then compute the gradient of the log-conditional explicitly:
\begin{align}
    \nabla_\theta\log G(\theta)(y\mid x)
    &=
    -\nabla_\theta E(x,y;\theta)
    \notag\\
    &\quad+
    \mathbb E_{y'\sim G(\theta)(\cdot\mid x)}
    \left[
        \nabla_\theta E(x,y';\theta)
    \right].
    \label{eq:ebm-score}
\end{align}
The expectation in the second term is over the output \(y'\) with the input \(x\) held fixed.

If gate \(G_j\) is of the specified form, using Eq.~\eqref{eqn:reinforce} gives
\begin{align}
\nabla_{\theta_j} J(\theta)
= \mathbb{E}_{(z^{(0)},\ldots,z^{(k)})}\Biggl[
  f\bigl(z^{(k)}\bigr)\Biggl(
    -\nabla_{\theta_j} E\bigl(z^{(j-1)},\, z^{(j)};\theta_j\bigr) \nonumber \\
    +\mathop{\mathbb{E}}_{z' \sim G_j(\cdot \mid z^{(j-1)})}
  \Bigl[\nabla_{\theta_j} E\bigl(z^{(j-1)},\, z';\theta_j\bigr)\Bigr]\Biggr)\Biggr]
\label{eq:ebm-gradient}
\end{align}

Thus, to estimate the gradient, one runs the full circuit to obtain the latent trajectory \((z^{(0)},\ldots,z^{(k)})\) and draws an independent auxiliary sample
\begin{equation*}
    z^{(j)\prime}
    \sim
    G_j(\theta_j)(\cdot\mid z^{(j-1)})   
\end{equation*}
to estimate the second term. The trajectory sample \(z^{(j)}\) can be reused for the first term, so only gate \(G_j\), clamped at the same input \(z^{(j-1)}\), must be run again. This estimator is inexpensive but, as a score-function estimator, can suffer from high
variance~\cite{greensmith2004variance}.

\section{Torx framework}\label{sec:torx-impl}

\subsection{Execution Interface}
\label{subsec:torx-interface}

\texttt{torx} is a JAX-based library~\cite{jax2018github,kidger2021equinox} for constructing, executing, and training PSCs. Circuits are defined by specifying a certain combination of kernels. A circuit is then lowered to its underlying directed factor graph, a DAG of kernels each exposing a \emph{sample} method. Given input values for a kernel, \emph{sample} returns output values drawn from the kernel's conditional distribution. Training is implemented following the hybrid loop of Figure~\ref{fig:hybrid-loop}. The circuit is executed to produce samples \(z^{(k)}\sim\rho_{\mathrm{out}}(\theta)\), which are used to estimate the objective \(J(\theta)\) of Eq.~\eqref{eq:readout-objective}, and the gradient estimators of Section~\ref{subsec:diff} supply the parameters update for the kernels, with JAX providing the automatic differentiation and compilation machinery.

\subsection{Backends}\label{sec:hard-disc}

Once a logical PSC has been constructed, one has to choose a \emph{backend} which specifies how the \emph{sample} calls of a circuit are executed. \texttt{torx} provides three software backends. \texttt{SampleSimulator} draws a sample from \(\rho_{\mathrm{out}}\) by sampling sequentially through the gates, and \texttt{HybridSampleSimulator} mirrors it for registers that mix discrete and continuous wires. \texttt{StateVectorSimulator} instead propagates the full probability vector, returning \(\rho_{\mathrm{out}}\) exactly up to floating-point error. Since the vector has size \(2^n\) for \(n\) pbits, it is primarily useful for small-scale prototyping.

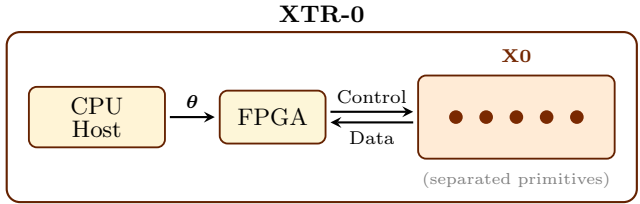
\begin{figure}[h]
\centering
\begin{tikzpicture}[
  >=stealth,
  line width=0.7pt,
  font=\small,
  host/.style   = {draw=extCopper!70!black, rounded corners=3pt, fill=extSolar!75,
                   minimum width=1.8cm, minimum height=0.8cm,
                   align=center, inner sep=3pt},
  fpganode/.style = {draw=extCopper!70!black, rounded corners=3pt, fill=extGold!22,
                   minimum width=1.4cm, minimum height=0.7cm,
                   align=center, inner sep=3pt},
  chipbox/.style = {draw=extCopper!70!black, rounded corners=3pt, fill=extOrange!16,
                   minimum width=2.6cm, minimum height=1.1cm,
                   align=center, inner sep=3pt},
  outerbox/.style = {draw=extCopper!70!black, rounded corners=5pt, thick,
                     inner sep=10pt},
  arr/.style    = {->, thick, shorten >=1.5pt, shorten <=1.5pt},
]

\node[host] (host) at (0, 0) {CPU\\[-2pt]Host};
\node[fpganode, right=0.7cm of host] (fp) {FPGA};
\node[chipbox,  right=1.2cm of fp]  (prim) {};

\node[font=\scriptsize\bfseries\color{extCopper!80!black}]
     at ($(prim.north)+(0,0.25)$) {X0};

\foreach \dx in {-0.8, -0.4, 0.0, 0.4, 0.8}{
  \fill[extCopper!85!black] ($(prim.center)+(\dx, 0)$) circle(0.09cm);
}

\node[font=\fontsize{6}{7}\selectfont, anchor=north, text=gray]
     at ($(prim.south)+(0,-0.05)$) {(separated primitives)};

\draw[arr] (host.east) -- node[above, font=\scriptsize]{\(\btheta\)} (fp.west);
\draw[arr] ([yshift= 2pt]fp.east) --
     node[above, font=\scriptsize]{Control} ([yshift= 2pt]prim.west);
\draw[arr] ([yshift=-2pt]prim.west) --
     node[below, font=\scriptsize]{Data}    ([yshift=-2pt]fp.east);

\node[outerbox,
      fit=(host)(fp)(prim),
      inner ysep=16pt, inner xsep=8pt,
      label={[font=\small\bfseries, anchor=south]above:XTR-0}] {};

\end{tikzpicture}
\caption{The XTR-0 desktop platform. A CPU host and FPGA
control physically separated pbit, pdit, Gaussian, and
mixture-of-Gaussian primitives on the socketed X0 chip. The FPGA
mediates data and control between primitives, so the logical
interaction topology and schedule are software-defined. The outer box
marks the complete XTR-0 platform.}
\label{fig:xtr0-arch}
\end{figure}

A hardware backend in \texttt{torx} replaces the software pseudorandom generator (e.g.~Threefry) with a physical randomness source. Any gate that needs a random draw can source it directly in hardware. The sigmoid-mixture gates of Section~\ref{sec:gates} illustrate this concretely. A gate of the form of Eq.~\eqref{eq:sigmoid-gate} is realized by programming a pbit (or pdit) to the mixture weight, then routing the register's values according to its outcome.

Concretely, we obtain randomness from X0, a subthreshold CMOS test chip accessed through the XTR-0 desktop platform, carrying pbits, pdits, Gaussian samplers, and mixture-of-Gaussian samplers as physically separated primitives. The host and FPGA set their parameters and route samples and control, so the logical interaction topology is software-defined (Figure~\ref{fig:xtr0-arch}). Operating transistors in the subthreshold regime exposes thermal and shot noise that X0 pbits use as a physical Bernoulli source, at energy costs of hundreds of attojoules to single femtojoules per sample at rates of tens of MHz~\cite{freitas2026taming,jelinvcivc2025efficient}, comparable to superparamagnetic tunnel junctions~\cite{parks2018superparamagnetic,vodenicarevic2018circuit,daniels2020energy}, while leveraging CMOS only process steps. However, the output statistics of such devices drift from device to device. For this reason, before sampling for the first time each pbit undergoes a Bayesian-optimization-based calibration that maps digital controls to bias voltage, with typical per-site calibration errors around 5\% in absolute probability. After calibration, a user programs target probabilities \(p\in[0,1]\) directly and receives binary outputs at rates set by the device relaxation time (\({\sim}100\;\mathrm{ns}\)). Appendix~\ref{app:x0-hardware} details the X0 probabilistic primitives and the XTR-0 platform.

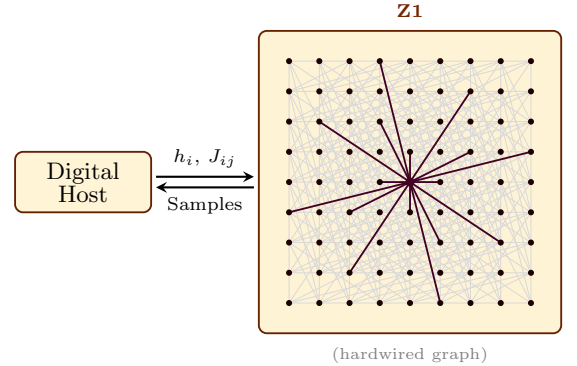
\begin{figure}[h]
\centering
\begin{tikzpicture}[
  >=stealth,
  line width=0.7pt,
  font=\small,
  host/.style   = {draw=extCopper!70!black, rounded corners=3pt, fill=extSolar!75,
                   minimum width=1.8cm, minimum height=0.8cm,
                   align=center, inner sep=3pt},
  chipbox/.style = {draw=extCopper!70!black, rounded corners=3pt, fill=extYellow!16,
                   minimum width=4.0cm, minimum height=4.0cm,
                   align=center, inner sep=3pt},
  outerbox/.style = {draw=extCopper!70!black, rounded corners=5pt, thick,
                     inner sep=10pt},
  arr/.style    = {->, thick, shorten >=1.5pt, shorten <=1.5pt},
]
\node[host] (host) at (0, 0) {Digital\\[-2pt]Host};
\node[chipbox, right=1.4cm of host] (z1chip) {};
\node[font=\scriptsize\bfseries\color{extCopper!80!black}]
     at ($(z1chip.north)+(0,0.25)$) {Z1};
\pgfmathsetmacro\gs{0.40}
\pgfmathsetmacro\Nmax{8}   
\begin{scope}[gray!30, line width=0.3pt]
  \foreach \ix in {0,...,\Nmax}{
    \foreach \iy in {0,...,\Nmax}{
      \pgfmathsetmacro\cx{(\ix - 4)*\gs}
      \pgfmathsetmacro\cy{(\iy - 4)*\gs}
      \foreach \dx/\dy in {1/0, 0/1, -1/2, 2/1, -3/2, 2/3, -1/4, 4/1}{
        \pgfmathtruncatemacro\jxi{\ix+\dx}
        \pgfmathtruncatemacro\jyi{\iy+\dy}
        \ifnum\jxi>-1 \ifnum\jxi<9 \ifnum\jyi>-1 \ifnum\jyi<9
          \pgfmathsetmacro\nx{(\jxi - 4)*\gs}
          \pgfmathsetmacro\ny{(\jyi - 4)*\gs}
          \draw ($(z1chip.center)+(\cx cm,\cy cm)$)
             -- ($(z1chip.center)+(\nx cm,\ny cm)$);
        \fi\fi\fi\fi
      }
    }
  }
\end{scope}
\begin{scope}[extFuchsia!85!black, line width=0.7pt]
  \foreach \dx/\dy in {
     1/0,  0/1,  -1/2,   2/1,  -3/2,   2/3,  -1/4,  4/1,
    -1/0,  0/-1,  1/-2,  -2/-1,  3/-2,  -2/-3,  1/-4, -4/-1}{
    \pgfmathsetmacro\tx{\dx*\gs}
    \pgfmathsetmacro\ty{\dy*\gs}
    \draw ($(z1chip.center)+(0,0)$)
       -- ($(z1chip.center)+(\tx cm,\ty cm)$);
  }
\end{scope}
\foreach \ix in {0,...,\Nmax}{
  \foreach \iy in {0,...,\Nmax}{
    \pgfmathsetmacro\cx{(\ix - 4)*\gs}
    \pgfmathsetmacro\cy{(\iy - 4)*\gs}
    \fill[extBrown] ($(z1chip.center)+(\cx cm,\cy cm)$) circle(0.04cm);
  }
}
\fill[extFuchsia!85!black] ($(z1chip.center)+(0,0)$) circle(0.06cm);
\node[font=\fontsize{6}{7}\selectfont, anchor=north, text=gray]
     at ($(z1chip.south)+(0,-0.05)$) {(hardwired graph)};
\draw[arr] ([yshift= 2pt]host.east) --
     node[above, font=\scriptsize]{\(h_i,\,J_{ij}\)}
     ([yshift= 2pt]z1chip.west);
\draw[arr] ([yshift=-2pt]z1chip.west) --
     node[below, font=\scriptsize]{Samples}
     ([yshift=-2pt]host.east);
\end{tikzpicture}
\caption{Z1 architecture overview. The digital host communicates bias fields \(h_i\) and couplings \(J_{ij}\) directly to the Z1 chip for Gibbs sampling. The chip hardwires a degree-16 interaction graph in silicon, each pbit (black dot) is physically coupled to 16 neighbors via on-die wiring that extends beyond nearest-neighbor positions. Fuchsia edges highlight one node's full neighborhood.}
\label{fig:z1-arch}
\end{figure}

Our next generation chip, which we refer to as Z1, hardwires a degree-16 sparse interaction graph directly in silicon and executes chromatic Gibbs sweeps without FPGA mediation, communicating only \(h_i\) and \(J_{ij}\) between sweeps (Figure~\ref{fig:z1-arch},~\cite{jelinvcivc2025efficient}). The two designs expose the substrate locality of Section~\ref{sec:stoc_circs} in opposite ways. XTR-0 reconstructs any \(k\)-local generator off-chip via the FPGA, paying a per-gate round-trip latency for arbitrary topology, while Z1 evaluates \(2\)-local Glauber generators natively in a single sweep but requires compilation of the logical PSC to a hardware-compatible representation~\cite{amico2026thermalizing}. The same logical PSC compiles to different gate schedules on each chip, and \texttt{torx} keeps the support \(I_j\) of every gate visible so that this compilation is explicit. Section~\ref{sec:hw-experiment} tests whether the X0 primitives provide Bernoulli randomness with statistics sufficient for the sampling algorithms of Section~\ref{sec:examples}.

\section{Demonstrations}\label{sec:examples}

We demonstrate the \texttt{torx} framework on a series of example sized to showcase the capabilities that a PSC can express rather than to show an efficiency or scalability advantage over standard implementations. Executable notebooks, including larger-scale variants of several of these examples, are available in the \texttt{torx} documentation~\cite{torxlib}, and we encourage readers to experiment with the framework directly.

The six demonstrations are: diffusion on a weighted graph (§\ref{sec:graph-diff}); discrete diffusion for image generation (§\ref{sec:discrete-diffusion}); stochastic graph networks (§\ref{sec:sgnn}); a regime-switching jump-diffusion process (§\ref{sec:jump-diffusion}); sampling and training an Ising model on a spin ring (§\ref{sec:ebm-sampling}); and the same Ising sampling task driven by hardware randomness (§\ref{sec:hw-experiment}). All demonstrations use software pseudorandomness except the last, which sources its randomness from calibrated pbits on X0.

\subsection{Random Walks on Graphs}\label{sec:graph-diff}
Consider an undirected weighted graph \(\mathcal{G}=(V,E)\) with edge weights \(w_{ij}=w_{ji}\geq 0\).
The graph Laplacian \(L=D-A\) satisfies \(L\mathbf{1}=0\), so \(Q=-L\) is a valid continuous-time Markov generator and the distribution \(x(t)\) over vertices satisfies \(\dot{x}=Qx\) with exact solution \(x(t)=e^{Qt}x(0)\).

We can write the \emph{edge decomposition}
\begin{equation}\label{eq:rw-decomp}
    Q = \sum_{\{i,j\}\in E} Q_{ij},\qquad Q_{ij} = -w_{ij}(e_i-e_j)(e_i-e_j)^\top,
\end{equation}
where each \(Q_{ij}\) is a local operator acting only on coordinates \(i\) and \(j\).
Since \(\gname{PSWAP}_{ij}(p) = I - p(e_i-e_j)(e_i-e_j)^\top\), setting \(p = w_{ij}\tau\) gives
\begin{equation}
    \gname{PSWAP}_{ij}(w_{ij}\tau) = I + \tau Q_{ij}.
\end{equation}
The exact exponential corresponds to \(\gname{PSWAP}_{ij}\!\left(\tfrac{1-e^{-2w_{ij}\tau}}{2}\right) = e^{\tau Q_{ij}}\).
Since \(Q_{ij}\) and \(Q_{kl}\) commute whenever edges \(\{i,j\}\) and \(\{k,l\}\) are vertex-disjoint (they act on disjoint coordinates), an edge coloring of \(\mathcal{G}\) groups commuting generators together, reducing the number of Trotter groups to \(k=\chi'(\mathcal{G})\) the edge chromatic number.

Applying Trotterization (Section~\ref{sec:trotterization}) with Euler approximations \(I + Q_{ij}t/N\), \(k=\chi'(\mathcal{G})\) local generators and \(N\) steps gives a global error
\begin{equation}\label{eq:rw-euler-trotter-error}
    \Bigl\|e^{Qt} - \Bigl(\prod_{\{i,j\}\in E} \bigl(I + Q_{ij}t/N\bigr)\Bigr)^{\!N}\Bigr\| = O\!\left(\frac{k^2 t^2}{N}\right),
\end{equation}
with each factor \(I + Q_{ij}t/N\) implemented as a single \(\gname{PSWAP}_{ij}\) gate.

While applying Trotterization with exact exponential \(e^{t/N Q_{ij}}\), \(k\) local generators and \(N\) steps gives a global error
\begin{equation}\label{eq:rw-trotter-error}
    \Bigl\|e^{Qt} - \Bigl(\prod_{\{i,j\}\in E} e^{Q_{ij}t/N}\Bigr)^{\!N}\Bigr\| = O\!\left(\frac{k^2 t^2}{N}\right),
\end{equation}
with each factor \(e^{Q_{ij}t/N}\) implemented as a single \(\gname{PSWAP}_{ij}\) gate.
Using Strang splitting gives an error of \(O(k^3 t^3/N^2)\) at the cost of \(2k-1\) gates per step rather than \(k\).

\begin{figure}[h]
    \centering
    \includegraphics[width=0.55\columnwidth]{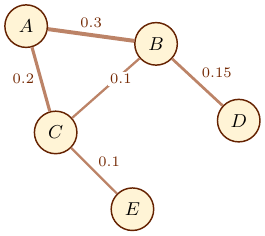}
    \caption{Weighted graph used for the diffusion example.}
    \label{fig:graph-and-laplacian}
\end{figure}

As an example, consider the graph in Figure~\ref{fig:graph-and-laplacian}.
 
\begin{figure}[h]
  \centering
\begin{tikzpicture}[
  >=stealth,
  font=\small,
  pwire/.style={thick, extBrown},
  swapgate/.style={draw, thick, fill=extGold!22, draw=extBrown,
                   rounded corners=2pt, inner sep=1pt, minimum width=0.4cm},
  wlbl/.style={font=\footnotesize, anchor=east, text=black!75},
  repeatbox/.style={draw=black!55, dashed, rounded corners=2pt, inner sep=3pt},
  rlbl/.style={font=\footnotesize\itshape, text=black!65, anchor=north},
]
\foreach \i/\lbl in {0/A, 1/B, 2/C, 3/D, 4/E} {
  \pgfmathsetmacro{\y}{-\i*0.55}
  \draw[pwire] (-0.4, \y) -- (5.6, \y);
  \node[wlbl] at (-0.45, \y) {\(\pket{\lbl}\)};
}

\foreach \x/\top/\bot in {0.4/0/1, 1.4/1/2, 2.4/0/2, 3.4/1/3, 4.4/2/4} {
  \pgfmathsetmacro{\ytop}{-\top*0.55}
  \pgfmathsetmacro{\ybot}{-\bot*0.55}
  \pgfmathsetmacro{\ymid}{(\ytop+\ybot)/2}
  \pgfmathsetmacro{\hgt}{abs(\ytop-\ybot)+0.35}
  \node[swapgate, minimum height=\hgt cm] at (\x, \ymid) {\(\gname{S}\)};
}
\node[repeatbox, fit={(0.1, 0.25) (4.7, -2.45)}] (rbox) {};
\node[rlbl] at (rbox.south) {\(\times\, N\)};
\end{tikzpicture}
  \caption{Single step of Trotterized circuit for graph diffusion. Each box, labeled \(\gname{S}\), is a \(\gname{PSWAP}\) gate acting on the pair of pbits it spans with probability proportional to the inverse number of Trotter steps; the dashed block is repeated \(N\) times.}
  \label{fig:diff-cir}
\end{figure}
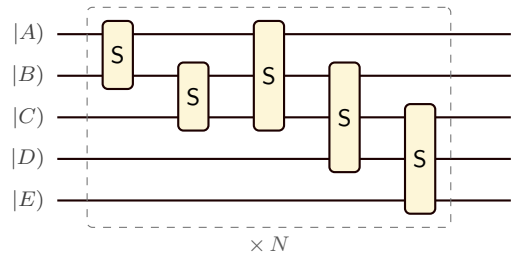

\begin{figure}[H]
    \centering
    \includegraphics[width=0.46\textwidth]{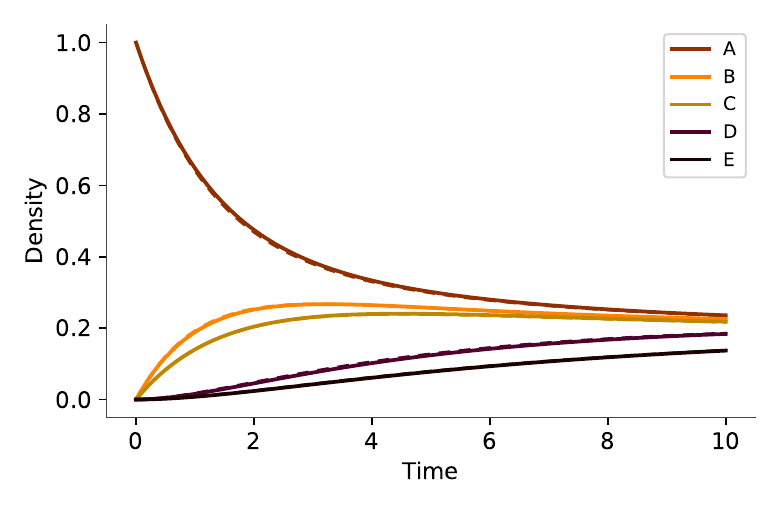}
    \caption{Long term graph diffusion trajectory. Trotter with Euler discretization is shown in dashed markings. Exact solution is shown in solid lines of the same color.}
    \label{fig:graph-diffusion-convergence}
\end{figure}

\paragraph{Simulation.}
Taking initial conditions \(x(0) = (1, 0, 0, 0, 0)\), corresponding to unit mass on state \(A\), we obtain a circuit consisting of a single walk layer of the form shown in Figure~\ref{fig:diff-cir}. 
Applying Trotterization with Euler discretization and iterating up to \(t = 10\), the resulting evolution of \(x(t)\) is plotted in Figure~\ref{fig:graph-diffusion-convergence}.

\subsection{Discrete Diffusion}\label{sec:discrete-diffusion}

Discrete diffusion models~\cite{austin2021structured,wang2024diffusion,sahoo2024simple,nie2025large} corrupt data with a forward Markov Chain and learn a neural network to reverse it.
Tau-leaping~\cite{gillespie2001tau} is an approximate simulation method that advances a jump process by a fixed time interval, treating many small events within that interval as independent.
We show that the tau-leaping reverse step can be implemented as a PSC of independent \(\gname{PNOT}\) gates, and that adding a small number of \(\gname{PCNOT}\) gates restores local pixel correlations that tau-leaping discards.

For the forward process, each bit \(x_i\in\{0,1\}\) evolves independently under the uniform rate matrix \(Q=\bigl(\begin{smallmatrix}-1&1\\1&-1\end{smallmatrix}\bigr)\), giving transition kernel
\begin{equation}
  e^{\sigma Q} =
  \begin{pmatrix}
    \tfrac{1}{2}+\tfrac{1}{2}e^{-2\sigma} & \tfrac{1}{2}-\tfrac{1}{2}e^{-2\sigma}\\[4pt]
    \tfrac{1}{2}-\tfrac{1}{2}e^{-2\sigma} & \tfrac{1}{2}+\tfrac{1}{2}e^{-2\sigma}
  \end{pmatrix}.
\end{equation}
The off-diagonal entry \(\tfrac{1}{2}(1-e^{-2\sigma})\) is the probability that a bit has flipped by noise level \(\sigma\).

A convolutional denoiser~\cite{ronneberger2015u} trained with binary cross-entropy produces per-pixel estimates \(\hat{p}_i = P(x_i^{(0)}=1\mid x_t)\).
The per-pixel flip probability for one reverse step is
\begin{equation}
  p^{(i)}_{\mathrm{flip}} =
  \begin{cases}
    1-\hat{p}_i & x_i = 1,\\
    \hat{p}_i   & x_i = 0,
  \end{cases}
\end{equation}
and the corresponding \(\gname{PNOT}\) logit is \(\theta^{(i)} = \log\bigl(p^{(i)}_{\mathrm{flip}}/(1-p^{(i)}_{\mathrm{flip}})\bigr)\).
The tau-leaping reverse step is therefore exactly the PSC, \(\bigotimes_i \gname{PNOT}(\theta^{(i)})\). 
It is composed of one independent gate per pixel, parametrized by the denoiser output.

Because tau-leaping treats all bits as independent, it misses joint structure.
For each pixel pair \((i,j)\) we compute the coupling
\begin{equation}\label{eq:dd-coupling}
  c_{j\to i} = p^{(i)}_{\mathrm{flip}}(x_t\mid\bar{\jmath}) - p^{(i)}_{\mathrm{flip}}(x_t),
\end{equation}
where \(\bar{\jmath}\) denotes \(x_t\) with bit \(j\) flipped, measuring how much \(j\)'s state shifts \(i\)'s flip probability.
We select the top \(K=50\) pairs with the maximum \(|c_{j\to i}|\) and append one \(\gname{PCNOT}(c_{j\to i})\) per pair, giving a circuit of \(n+K\) gates in total.

\begin{figure}[h]
\centering
\includegraphics[width=1.0\linewidth]{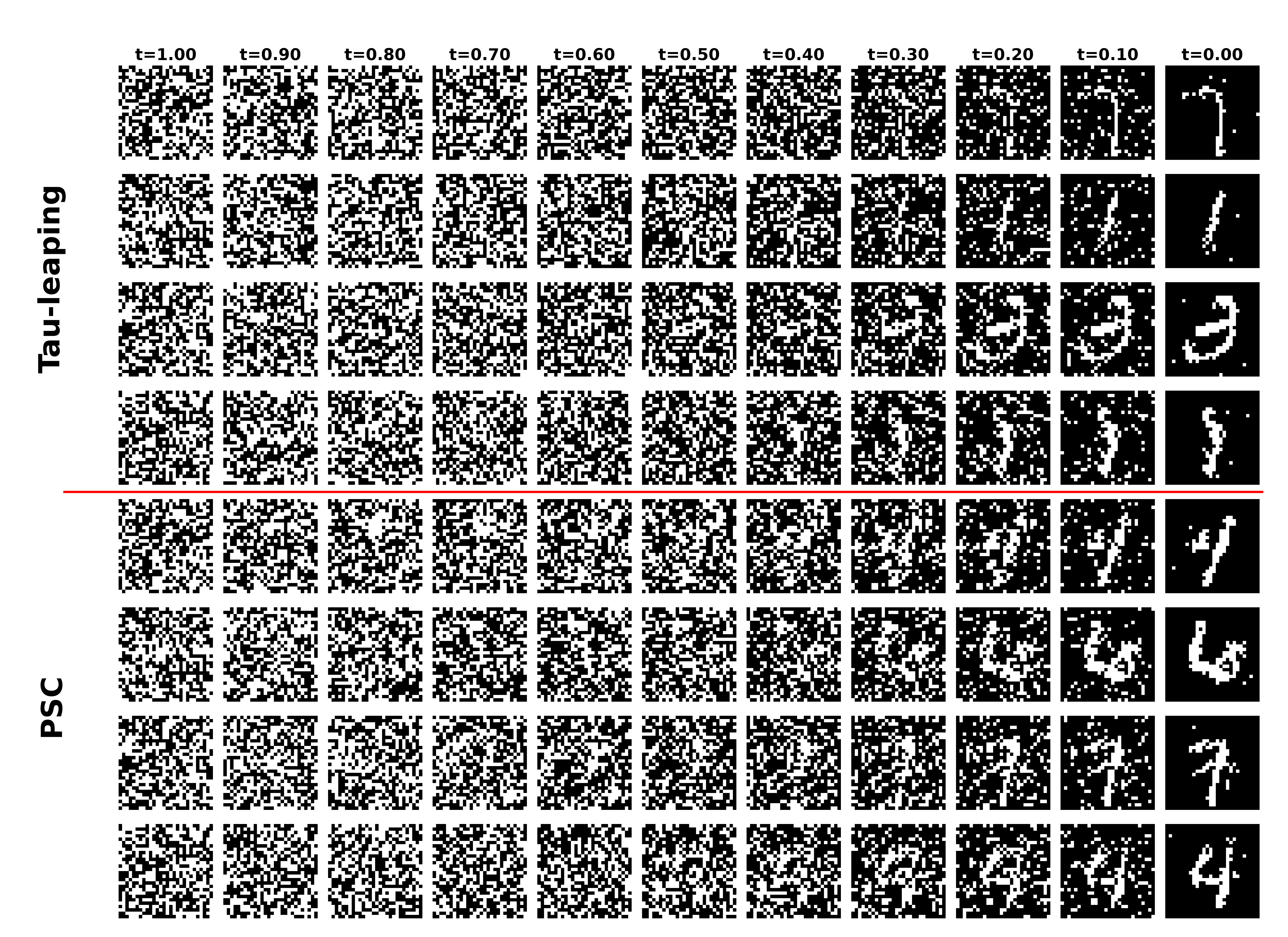}
\caption{A comparison of resulting images from the tau-leaping inference method and the augmented PSC method. We show the results for 10 solver steps.}
\label{fig:discrete-diff}
\end{figure}

Results on \(28\times28\) binary MNIST are shown in Figure~\ref{fig:discrete-diff}: the PSC augmentation reduces bit-error rate from \(0.114\) (tau-leaping) to \(0.113\) and closes \(41\%\) of the FID gap between corrupted and clean distributions, demonstrating that denoiser outputs can parameterize structured PSC proposals.

\subsection{Stochastic Graph Networks}\label{sec:sgnn}

\emph{Stochastic geometric deep learning}~\cite{bronstein2021geometric,cohen2016group} builds symmetry into a stochastic model rather than learning it from data.
The model is constrained to be equivariant under a group \(G\) of transformations of the input.
In our implementation of a stochastic graph network as a PSC this is realized by parameter sharing where the same local gate kernel is reused wherever a group element relates two regions of the wire layout.

A \emph{Stochastic Graph Network} (SGNN) instantiates this for the symmetry group of graph automorphisms~\cite{verdon2019quantum,zhou2020graph}: one wire per node, one parametrized \(\gname{PIsing}\) gate per edge, with parameters tied under any automorphism that permutes the edges.
Assigning pbit \(\sigma_i\in\{0,1\}\) to node \(i\) with spin \(s_i=2\sigma_i-1\in\{-1,+1\}\), the gate for edge \((i,j)\) is
\begin{equation}\label{eq:sgnn-gate}
  G_{ij}(\theta_{ij}) = \gname{PIsing}(J_{ij},\,h_i,\,h_j,\,\beta,\,\Delta t),
\end{equation}
where \(J_{ij}\in\mathbb{R}\) is a learnable edge coupling and \((\beta,\Delta t)\) are fixed hyperparameters.
One \emph{sweep} is a series of gate layers acting on a color partition $k$ of the graph,
\begin{equation}\label{eq:sgnn-layer}
  \widetilde{L}(\theta)
  = \Bigl(\bigotimes_{k} G_{k}(\theta)\Bigr)\otimes\Id_{\setminus k},
\end{equation}
and the full SGNN circuit repeats the sweep \(N\) times.

\begin{figure}
  \centering
  \includegraphics[width=0.76\linewidth]{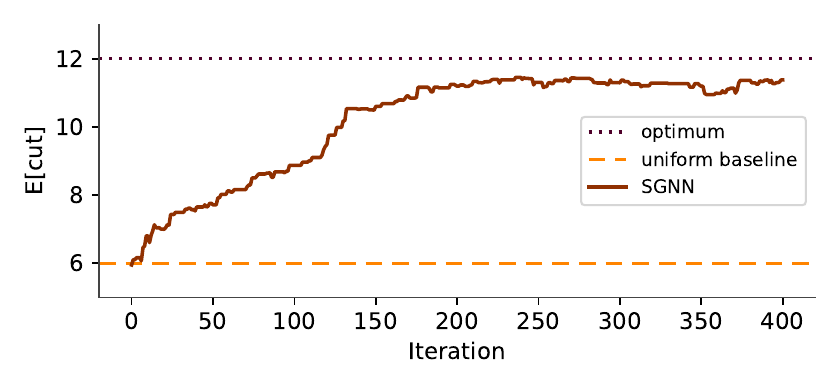}
  \caption{Expected cut value during SGNN training on a random 3-regular 8-site graph (\(|E|=12\), optimal cut \(=12\)).
           One \(\gname{PIsing}\) gate per edge, \(N=5\) sweeps, \(S=1024\) samples per step.
           Dashed line: uniform random baseline; dotted line: optimum.}
  \label{fig:sgnn-maxcut}
\end{figure}

Writing \(\mathrm{cut}(\bm\sigma)=\sum_{(i,j)\in E}(1-s_is_j)/2\), the objective is to maximize the expected cut under the circuit output \(\rho_\theta=\graph_{\mathrm{SGNN}}(\theta)\,\rho_0\):
\begin{equation}\label{eq:sgnn-loss}
  \mathcal{L}(\theta)
  = -\Ex_{\rho_\theta}[\mathrm{cut}]
  = -\sum_{(i,j)\in E}\frac{1-\langle s_is_j\rangle_{\rho_\theta}}{2}.
\end{equation}
The circuit runs in sample mode with \(\mathcal{L}\) estimated by Monte Carlo and edge couplings are optimized by a Metropolis search, which is one of several applicable optimization strategies; gradient-based alternatives such as REINFORCE or the parameter-shift rule of Section~\ref{subsec:diff} are equally applicable.
At each step a Gaussian perturbation is proposed to \(\theta\), \(S\) samples are drawn from the updated circuit, the average cut is estimated, and the proposal is accepted with probability \(\min(1, e^{-\Delta\mathcal{L}/T})\). 
On a random 3-regular graph (\(n=8\), \(|E|=12\)) with one \(\gname{PIsing}\) gate per edge, \(N=5\) sweeps, and \(S=1024\) samples per step, the trained SGNN outperforms a uniform sampler and recovers the optimal cut in the majority of runs (Figure~\ref{fig:sgnn-maxcut}).
For context, the Goemans--Williamson SDP relaxation~\cite{goemans1995improved} guarantees a \(0.878\)-approximation in the worst case over all graphs; this is a different kind of claim than the instance-specific result reported here, where the trained SGNN matches the optimal cut on this graph but carries no worst-case guarantee.

\subsection{Jump Diffusion}\label{sec:jump-diffusion}

Regime-switching models are a classical family of hybrid Markov processes in which a latent discrete state modulates the dynamics of a continuous variable~\cite{hamilton1989new}. The joint state \((S_t, X_t) \in \{0,\ldots,K{-}1\} \times \R\) evolves under
\begin{equation}\label{eq:rs_sde}
    \dd X_t = \mu_{S_t}\,\dd t + \sigma_{S_t}\,\dd W_t,
\end{equation}
where \(S_t\) is a continuous-time Markov chain, and the drift \(\mu_k\) and volatility \(\sigma_k\) are conditioned on the discrete state, as in Eq.~\eqref{eq:mog-mixture}. This structure results in what is effectively a hidden Markov model (HMM) in continuous time, the observed process \(X_t\) is driven by a latent chain \(S_t\). 

The generator of the full continuous time process decomposes as \(A = A_S + A_X\), where \(A_S\) governs discrete transitions and \(A_X\) generates the conditioned diffusion. These terms do not commute because the drift and volatility in \(A_X\) depend on the discrete state \(S\). Writing \(T_G(\Delta t)\) for the finite-\(\Delta t\) gate implementing generator \(G\), we split the one-step propagator into two local gates:
\begin{equation}\label{eq:rs_trotter_step}
    T_{\mathrm{Trotter}}(\Delta t) = T_{\gname{MoG}}(\Delta t)T_{\gname{PditCycle}}(\Delta t).
\end{equation}
The first factor discretizes the chain. We use a \(\gname{PditCycle}\) gate on a \(K\)-state wire, which implements a cyclic random walk with forward rate \(\lambda_+\) and backward rate \(\lambda_-\) which determine the probability of moving forward, cycling backward, or staying. in the current position \(p_\pm = \lambda_\pm \Delta t\), \(p_0 = 1 - (\lambda_+ + \lambda_-)\Delta t\). This results in an Euler discretization as we have seen before.

\begin{figure}
  \centering
\begin{tikzpicture}[
  >=stealth,
  font=\small,
  pditwire/.style={thick, extOrange!80!black, double, double distance=1.6pt},
  pmodewire/.style={thick, extFuchsia, decorate,
                    decoration={snake, amplitude=0.5mm, segment length=2.5mm}},
  pditgate/.style={draw, thick, fill=extOrange!18, draw=extOrange!80!black,
                   rounded corners=2pt, inner sep=2pt},
  pmodegate/.style={draw, thick, fill=extFuchsia!22, draw=extFuchsia,
                    rounded corners=2pt, inner sep=2pt},
  ctrldot/.style={circle, fill=extOrange!80!black, draw=extOrange!80!black,
                  inner sep=0pt, minimum size=0.18cm},
  ctrllink/.style={thick, extOrange!50!extFuchsia},
  wlabel/.style={font=\footnotesize, anchor=east, text=black!75},
]
\draw[pditwire]  (-0.4, 0.55) -- (3.8, 0.55);
\draw[pmodewire] (-0.4,-0.55) -- (3.8,-0.55);

\node[wlabel] at (-0.5, 0.55) {\(\pket{S}\)};
\node[wlabel] at (-0.5,-0.55) {\(\pket{X}\)};

\node[pditgate,  minimum width=1.5cm, minimum height=0.5cm] at (1.0,  0.55) {\(\gname{PditCycle}\)};

\draw[ctrllink] (2.7, 0.55) -- (2.7, -0.55);
\node[ctrldot]  at (2.7, 0.55) {};
\node[pmodegate, minimum width=1.0cm, minimum height=0.5cm] at (2.7, -0.55) {\(\gname{MoG}\)};
\end{tikzpicture}
  \caption{One Trotter step for the circuit. The \(\gname{PditCycle}\) gate transitions the pdit, then the \(\gname{MoG}\) gate applies regime-conditioned Gaussian dynamics to the continuous wire. Repeating this unit \(N = T/\Delta t\) times yields the full circuit.}
  \label{fig:rs_circuit}
\end{figure}
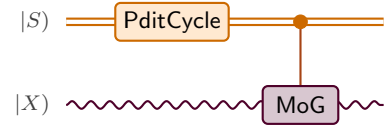

The second factor applies a \(\gname{MoG}\) gate, the hybrid gate of Eq.~\eqref{eq:mog-mixture}, conditioned on the current \(S = k\):
\begin{equation}\label{eq:rs_mog_step}
    X^\prime =  X + \mu_k\,\Delta t + \sigma_k\sqrt{\Delta t} \mathcal{N}(0,1).
\end{equation}
This is an Euler step for the diffusion in Eq.~\eqref{eq:rs_sde}. The full one-step circuit is shown in Figure~\ref{fig:rs_circuit}.

\begin{figure}
    \centering
    \includegraphics[width=0.86\columnwidth]{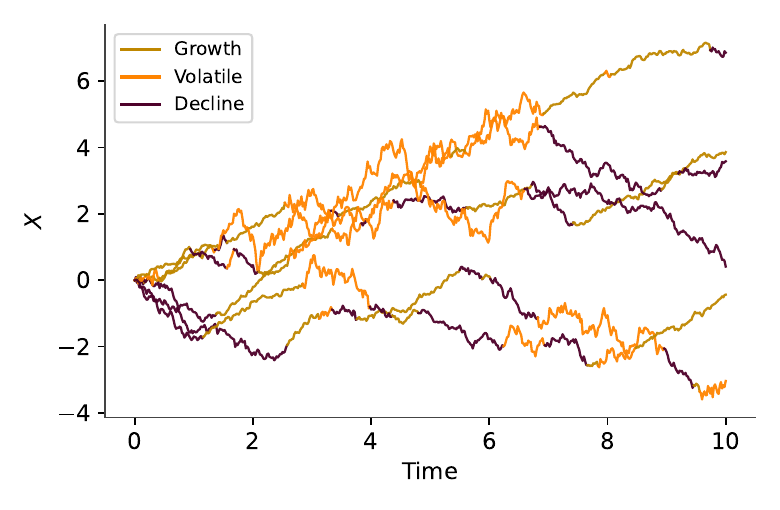}
    \caption{Sample trajectories from the jump diffusion circuit, colored by the latent at each step.}
    \label{fig:rs_trajectories}
\end{figure}

To test this example, we set \(K = 3\) regimes with drifts \(\mu = (1.0, 0.0, -0.8)\), volatilities \(\sigma = (0.3, 1.0, 0.5)\), and transition rates \(\lambda_+ = 0.5\), \(\lambda_- = 0.3\), simulating to time \(T = 10\). Figure~\ref{fig:rs_trajectories} shows sample trajectories colored by the latent regime at each step. Since \(A_S\) and \(A_X\) do not commute, the Trotterization incurs an error linear in \(\Delta t\), which can be reduced by decreasing the step size or moving to a higher-order method.

\subsection{Ising Sampling}\label{sec:ebm-sampling}

The Ising model assigns probability \(\pi(\bm{s})\propto e^{-\beta H(\bm{s})}\) to binary configurations via the Hamiltonian \(H(\bm{s}) = -\sum_{(i,j)\in E} J_{ij} s_i s_j - \sum_i h_i s_i\), \(s_i\in\{-1,+1\}\).
We demonstrate on an 8-site ring with random couplings \(J_{ij}\) and fields \(h_i\) at \(\beta=1.5\).

Since the ring is two-colorable, even-indexed sites \(\{0,2,4,6\}\) and odd-indexed sites \(\{1,3,5,7\}\) each form a conditionally independent class~\cite{gonzalez2011parallel}.
Writing \(\sigma_i=(s_i+1)/2\in\{0,1\}\) for the circuit variable, the Gibbs conditional at site \(i\) is
\begin{align}
  \ell_i &= h_i + \sum_{j\in N(i)} J_{ij}(2\sigma_j-1), \\
  \pi(\sigma_i=1\mid\sigma_{N(i)}) &= \frac{1}{1+e^{-2\beta\ell_i}}.
\end{align}
One sweep applies two layers in sequence; within each layer every site in that color class is updated in parallel by
\begin{equation}
  \gname{PColor}_i = \gname{PNOT}(2\beta\ell_i)\circ\gname{PReset}(\infty),
\end{equation}
which resets \(\sigma_i\) to \(0\) and resamples it from the exact Gibbs conditional.
For this we need to compute \(\ell_i\) on the fly at each sweep, which we do classically in Python, although on our Z1 chip this can be done natively~\cite{amico2026thermalizing}.
Over \(6000\) chains and \(240\) sweeps this achieves a total-variation distance of \(0.052\) from the exact Boltzmann distribution.

\begin{figure}[h]
    \centering
    \includegraphics[width=1.0\linewidth]{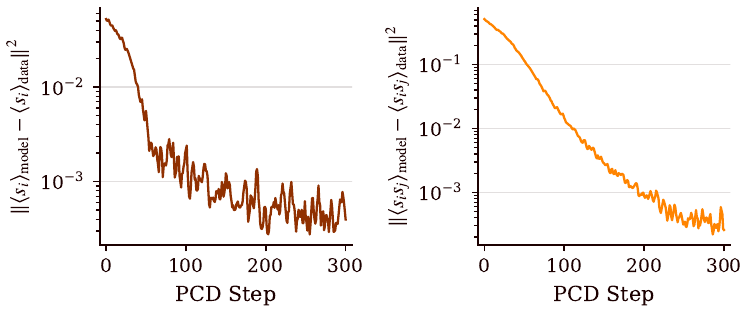}
    \caption{MSE of learned parameters relative to the true Boltzmann distribution during PCD training on the 8-site ring. Left: magnetization \(h_i\); right: pairwise couplings \(J_{ij}\).}
    \label{fig:chroma-gibbs-params}
\end{figure}

The parameters \((J_{ij},h_i)\) can be fit to data using Persistent Contrastive Divergence~\cite{tieleman2008training}, the special case of the energy-based-kernel gradient of Section~\ref{subsec:diff} in which the kernel samples from the Ising Gibbs conditional: \(2048\) persistent chains run \(2\) chromatic sweeps per step to estimate model expectations, giving the log-likelihood gradient
\begin{equation}\label{eq:pcd-grad}
  \frac{\partial \log \mathcal{L}}{\partial J_{ij}}
  = \beta\bigl(\langle s_i s_j\rangle_{\mathrm{data}} - \langle s_i s_j\rangle_{\mathrm{model}}\bigr),
\end{equation}
and analogously for \(h_i\).
Over 300 steps at learning rate \(0.08\) the learned parameters converge to the true couplings; see Figure~\ref{fig:chroma-gibbs-params}.

A natural alternative assigns one \(\gname{PIsing}\) gate per edge and Trotterizes their composition.
As \(N\to\infty\), this converges to the semigroup generated by \(\sum_{(i,j)\in E} Q_{ij}\), but this generator is not the Glauber generator for the full ring.
Each two-site generator \(Q_{ij}\) gives site \(i\) a flip rate that depends only on the single neighbor \(j\) via \(\sigma\!\bigl(-2\beta J_{ij}(2\sigma_j-1)\bigr)\); the effective single-site rate from composing all edge generators is \(\sum_{j\in N(i)}\sigma\!\bigl(-2\beta J_{ij}(2\sigma_j-1)\bigr)\).
The correct Glauber rate is \(\sigma(-2\beta\ell_i)\) with the full local field \(\ell_i=h_i+\sum_{j\in N(i)}J_{ij}(2\sigma_j-1)\) inside the sigmoid.
Since \(\sigma\) is nonlinear, \(\sum_j\sigma(\alpha_j)\neq\sigma(\sum_j\alpha_j)\) in general, so the stationary distribution of \(\sum_{\mathrm{edges}}Q_{ij}\) is not the Boltzmann distribution.
On the same 8-site ring this yields TV distance \(0.402\), nearly eight times worse than chromatic Gibbs.

\begin{figure}[h]
    \centering
    \includegraphics[width=1.0\linewidth]{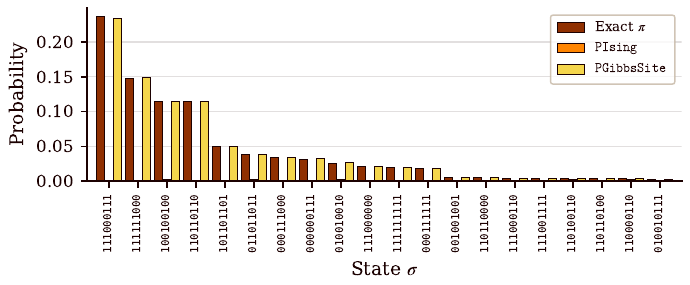}
    \caption{State probabilities on the 8-site ring comparing the exact Boltzmann distribution, edge-tiled \(\gname{PIsing}\), and the per-site Gibbs gate. States are ordered by exact probability.}
    \label{fig:gibbs-site-comparison}
\end{figure}

The exact Gibbs conditional can instead be embedded into a single gate acting on site \(i\) and all its \(d\) neighbors.
This \(2^{d+1}\times 2^{d+1}\) matrix stores \(\gname{PNOT}(2\beta\ell_i)\circ\gname{PReset}(\infty)\) for each of the \(2^d\) neighbor configurations statically, unlike \(\gname{PColor}_i\) whose \(\ell_i\) is recomputed dynamically at each sweep.
Figure~\ref{fig:gibbs-site-comparison} confirms that the per-site gate matches the exact distribution, while edge-tiled \(\gname{PIsing}\) does not.
The same gate applies directly to any Ising-type inference task, including Boltzmann Machines~\cite{ackley1985learning} and DTMs~\cite{jelinvcivc2025efficient}.

\subsection{Ising Sampling with Hardware Randomness}
\label{sec:hw-experiment}

\begin{figure*}[t]
  \centering
  \begin{subfigure}[t]{0.48\textwidth}
    \centering
    \includegraphics[width=\linewidth]{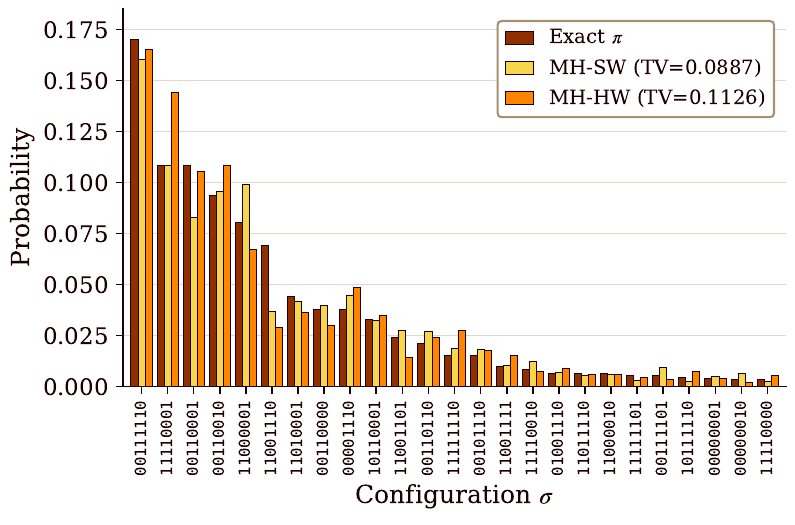}
    \caption{Sampling with a Metropolis-Hastings method.}
    \label{fig:hw-mh}
  \end{subfigure}\hfill
  \begin{subfigure}[t]{0.48\textwidth}
    \centering
    \includegraphics[width=\linewidth]{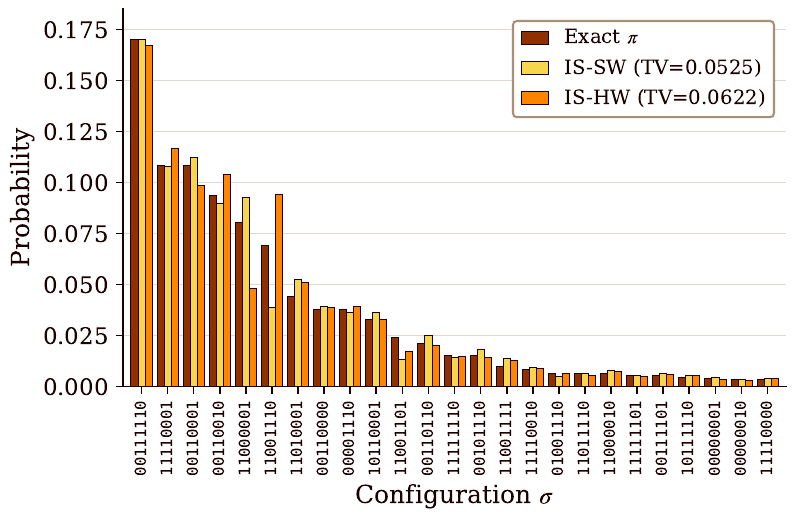}
    \caption{Sampling with importance reweighting.}
    \label{fig:hw-is}
  \end{subfigure}
  \caption{Experimental validation of sampling with noise from XTR-0. The exact distribution is shown in dark red, with NumPy-based pseudo-randomness in yellow, and hardware randomness in orange (MH and IS). The target distribution is an 8-site Ising ring (with \(\beta = 1.5\) and \(20{,}000\) samples) using a calibrated pbit on X0 within the XTR-0 platform.}
  \label{fig:hw-distributions}
\end{figure*}

We sample the stationary distribution \(\pi\) of the same 8-site Ising ring at
\(\beta=1.5\) of Section~\ref{sec:ebm-sampling} using Bernoulli randomness from a calibrated pbit on X0, accessed through XTR-0, with proposals \(q(\cdot;\theta)\) drawn as independent per-site Bernoulli samples at the pbits' programmable biases \(\theta_i\). Two algorithms are tested: Metropolis-Hastings (MH)~\cite{metropolis1953equation,hastings1970monte}, which uses hardware bits to inform the acceptance ratio \(a = \min\!\left(1,\tfrac{\pi(x')}{\pi(x)}\tfrac{q(x)}{q(x')}\right)\), and self-normalized importance sampling (IS)~\cite{kahn1951estimation,kloek1978bayesian}, which weights each sample by \(\tilde{w}(m)=\pi(x^{(m)})/q(x^{(m)};\theta)\). The exact distribution \(\pi\) is available by full enumeration over the \(2^8\) configurations. Both methods achieve a total variation distance consistent with software baselines, using 20,000 hardware-drawn samples (Figure~\ref{fig:hw-distributions}), demonstrating that analog hardware provides randomness of competitive quality to its digital counterpart.

\section{Conclusions}
\label{sec:discussion}

We have introduced Parametrized Stochastic Circuits as a gate-based model for programmable stochastic dynamics, together with \texttt{torx}~\cite{torxlib}, a differentiable JAX framework for building, executing, and training them. In this model, local stochastic kernels with tunable settings compose into a circuit operator, the gradient estimators of Section~\ref{subsec:diff} make
the circuit trainable from samples, and a backend specifies the execution rule. The interface supports exact probability-vector simulation, sample-based execution, and hardware-sourced randomness while preserving the same logical PSC object.

The gate library covers directly specified stochastic kernels, generator-derived kernels such as \(\gname{PIsing}\), and Trotterized approximations of local continuous-time Markov generators. The demonstrations use these kernels for random walks on graphs, discrete diffusion, stochastic graph networks, jump diffusion and Ising sampling. In the final demonstration, calibrated subthreshold-CMOS pbits on X0 supply Bernoulli randomness to the same Ising sampling task, achieving total-variation distances consistent with software baselines. The logical circuits are unchanged, and the only difference between them is the randomness source.

The companion manuscript~\cite{amico2026thermalizing} on the \texttt{thermalizers} framework introduces a complementary hardware route, in which clamping the input variables of an energy-based model and sampling its output variables realizes a stochastic kernel directly as physical dynamics. In this relation, \texttt{torx} specifies and trains the logical PSC kernel, while the \texttt{thermalizers} construction implements it natively in hardware. Beyond this, several directions remain open: executing PSCs natively on Z1, a general theory of synthesizing a target kernel from a fixed gate set, and scaling the demonstrations beyond the toy sizes considered here. We view the PSC framework and \texttt{torx} as a concrete starting point for programming probabilistic hardware.

\section{Acknowledgments}

The authors thank Jack Ceroni and Edward Jiang for their contributions to early prototypes of the software, the authors also thank Nahuel Freitas and Andraž Jelinčič for their valuable feedback.

\bibliography{apssamp,paper_edited_refs}

\clearpage

\appendix

\section{Additional Gates}\label{app:gates}

Here we list some additional elementary gates that one might consider (and are supported by \texttt{torx}).
We note that all of the following gates we implement can be written as,
\[
\gname{ApplyB}(p) = (1-p)\Id + pB,
\]
where \gname{ApplyB} is the name of the operation and it applies \(B\) with probability \(p\), otherwise does nothing.
Setting \(p\in\{0,1\}\) in the following gates leads to deterministic gate.

\paragraph{\gname{PReset}}
Sets the pbit to \(0\) with probability \(p\), otherwise does nothing.
\[
\gname{PReset}(p) = \begin{pmatrix} 1 & p \\ 0 & 1-p \end{pmatrix}
\]

\paragraph{\gname{PCNOT}}
Flips the second pbit if the first is \(1\), with probability \(p\).
\[
\gname{PCNOT}(p) = \begin{pmatrix} 1 & 0 & 0 & 0 \\ 0 & 1 & 0 & 0 \\ 0 & 0 & 1-p & p \\ 0 & 0 & p & 1-p \end{pmatrix}
\]

\paragraph{\gname{PSWAP}}
Swaps two pbits with probability \(p\), otherwise does nothing.
\[
\gname{PSWAP}(p) = \begin{pmatrix} 1 & 0 & 0 & 0 \\ 0 & 1-p & p & 0 \\ 0 & p & 1-p & 0 \\ 0 & 0 & 0 & 1 \end{pmatrix}
\]

\paragraph{\gname{PJUMP}}
Moves probability from \(|10)\) to \(|01)\) with probability \(p\).
\[
\gname{PJUMP}(p) = \begin{pmatrix} 1 & 0 & 0 & 0 \\ 0 & 1 & p & 0 \\ 0 & 0 & 1-p & 0 \\ 0 & 0 & 0 & 1 \end{pmatrix}
\]

\paragraph{\gname{PDEMUX}}
Copies first pbit to second pbit and resets input to \(0\) with probability \(p\), otherwise does nothing (a probabilistic demultiplexor).
\[
\gname{PDEMUX}(p) = \begin{pmatrix} 1 & p & 0 & 0 \\ 0 & 1-p & p & p \\ 0 & 0 & 1-p & 0 \\ 0 & 0 & 0 & 1-p \end{pmatrix}
\]

\paragraph{\gname{PCopy}}
Copies first pbit to second pbit with probability \(p\), otherwise does nothing.
\[
\gname{PCopy}(p) = \begin{pmatrix} 1 & p & 0 & 0 \\ 0 & 1-p & 0 & 0 \\ 0 & 0 & 1-p & 0 \\ 0 & 0 & p & 1 \end{pmatrix}
\]

\paragraph{\gname{PditShift}}
Shifts a \(k\)-dimensional pdit cyclically \(i \to (i-1) \bmod k\) with probability \(p\). For \(k=3\):
\[
\gname{PditShift}(p) = \begin{pmatrix} 1-p & p & 0 \\ 0 & 1-p & p \\ p & 0 & 1-p \end{pmatrix}
\]

\paragraph{\gname{PditSWAP}}
Swaps two \(k\)-dimensional pdits with probability \(p\). The \(k^2 \times k^2\) matrix permutes \((i,j) \leftrightarrow (j,i)\):
\[
\gname{PditSWAP}(p) = (1-p)\Id + pS
\]
where \(S\) is the swap permutation matrix.

\section{Guide for readers from Quantum Information}\label{app:qi-guide}
  
The visual similarity of PSCs to Parametrized Quantum Circuits (PQCs)~\cite{benedetti2019parameterized}, composition of basic gates to create circuits, hides several substantive distinctions in data encoding, circuit-run accounting, and gradient estimation. The variational loop of Figure~\ref{fig:hybrid-loop} is deliberately the stochastic analog of the parametrized-quantum-circuit training loops developed in the quantum machine learning literature~\cite{verdon2017quantumalgo,verdon2018universal,broughton2020tensorflowquantum}, where a parametrized unitary is sampled, a classical optimizer evaluates a scalar loss, and the circuit parameters are updated. Indeed, several of the constructions we use have direct precursors in that body of work: variational training of a parametrized circuit~\cite{verdon2017quantumalgo,verdon2018universal}, classical meta-optimizers that warm-start the circuit parameters~\cite{verdon2019metalearn}, symmetry-tied gates on a graph~\cite{verdon2019quantum,zhou2020graph}, and the preparation of thermal (Gibbs) states by a variational procedure~\cite{verdon2019thermalizer}. Software frameworks such as TensorFlow-Quantum~\cite{broughton2020tensorflowquantum} and PennyLane~\cite{bergholm2018pennylane} were built to express exactly such hybrid quantum-classical loops and to differentiate through them, and the PSC software model can be read as a stochastic-matrix counterpart to them. The remainder of this section makes the distinctions precise.

In quantum information, computations are encoded as unitary matrices, whereas in Parametrized Stochastic Circuits computations are encoded as stochastic matrices. These two classes of objects differ in fundamental ways, which constrain the sorts of computations that quantum and stochastic computers can perform. First, a stochastic matrix need not have an inverse, unlike a unitary. The deterministic reset of one bit to 0 is
  \begin{equation}
      K_{\mathrm{reset}}=\begin{pmatrix}1&1\\0&0\end{pmatrix},
      \qquad
      K_{\mathrm{reset}}\pket{0}=K_{\mathrm{reset}}\pket{1}=\pket{0};
  \end{equation}
which has rank 1 and therefore no inverse. Mixing, marginalization, and coarse-graining are likewise valid stochastic circuit operations that generally cannot be inverted. Second, the two models differ in how information is read out. The state of a quantum register is a vector of complex amplitudes, so a readout requires the choice of a measurement basis, and the measurement collapses the state, such that subsequent operations act on the post-measurement state rather than on the original one. In contrast, the state of a PSC register is a classical probability distribution, and a readout is simply a sample drawn from it. Drawing a sample does not alter the distribution, so one can take measurements at any point in the circuit without affecting the remainder of the computation.

\section{Pmode Gate Library and Constructions}\label{app:pmode}

This appendix collects the detailed gate library for (\textsf{pmode}) introduced in Section~\ref{sec:gates}.

\subsection{Gaussian Stochastic Maps}

A pmode transformation \(\mathcal T:\mathcal P(\R^N)\to\mathcal P(\R^N)\) is \emph{Gaussian} if it sends every Gaussian density to a Gaussian density. Equivalently, it is \emph{affine in moments}: there exist \(M\in\R^{N\times N}\), \(\bm d\in\R^N\), \(\Delta\in\mathrm{Sym}^{\geq 0}_N(\R)\) with
\begin{align}
    \bm\mu\longmapsto\bm\mu'=M\bm\mu+\bm d,
    \qquad
    \Sigma\longmapsto\Sigma'=M\Sigma M^\top+\Delta.
    \label{eq:gauss-moment-update}
\end{align}
Collecting the data of a Gaussian stochastic map as a triple \((M,\bm d,\Delta)\) and equipping it with the composition law
\begin{equation}
    \begin{aligned}
    (M_2,\bm d_2,\Delta_2)\circ(M_1,\bm d_1,\Delta_1)
    &=
    \bigl(
    M_2 M_1,\;
    M_2\bm d_1+\bm d_2,
    \\
    &\qquad
    M_2\Delta_1 M_2^\top+\Delta_2
    \bigr)
    \end{aligned}
    \label{eq:sn-comp}
\end{equation}
gives a semigroup of Gaussian stochastic maps, denoted here by \(\mathcal S_N\). Composition is associative, the identity is \((\Id,\bm 0,0)\), and \((M,\bm d,\Delta)\) has a two-sided inverse only when \(\Delta=0\) and \(M\) is invertible; any nonzero noise injection is irreversible at the level of Markov kernels. Classical stochastic systems live on configuration space alone, not on phase space, so the symplectic form does not appear and three independent ingredients, linear map \(A\), drift \(b\), and noise injection \(D\), replace the single Hamiltonian generator of the CV-quantum case.

The finite-time propagator of a linear SDE \(\dd X_t=(AX_t+b)\,\dd t+B\,\dd W_t\) is the one-parameter sub-semigroup of \(\mathcal S_N\) given by
\begin{align}
    M(t)&=e^{At},
    \nonumber\\
    \bm d(t)&=\int_0^t e^{A(t-s)}b\,\dd s,
    \nonumber\\
    \Delta(t)&=\int_0^t e^{As}\,D\,e^{A^\top s}\,\dd s,
    \label{eq:gauss-propagator}
\end{align}
with \(D=BB^\top\). The moments of a Gaussian initial condition obey \(\dot{\bm\mu}_t=A\bm\mu_t+b\) and the differential Lyapunov equation \(\dot\Sigma_t=A\Sigma_t+\Sigma_t A^\top+D\); Gaussian pmode calculations therefore reduce to finite-dimensional linear algebra.

\paragraph{SVD-based Gaussian-channel factorization.}
Any \((M,\bm d,\Delta)\in\mathcal S_N\) with invertible \(M\) factors as
\begin{align}
    (M,\bm d,\Delta)
    &=
    (\Id,\bm d_2,\Delta_2)\circ(U_2,\bm 0,0)
    \nonumber\\
    &\quad\circ
    (\Sigma_S,\bm 0,\Delta_S)
    \circ(V^\top,\bm 0,0)\circ(\Id,\bm d_1,0),
    \label{eq:stoch-blochmessiah}
\end{align}
where \(M=U_2\Sigma_S V^\top\) is the SVD with \(U_2,V\in\mathrm{O}(N)\), \(\Sigma_S=\mathrm{diag}(\sigma_1,\ldots,\sigma_N)\), \(\Delta_S\) is diagonal in the singular-vector basis, and \(\Delta_2\) absorbs residual cross-mode noise when such a decomposition of the noise covariance is available. Each orthogonal factor can be decomposed into planar two-mode rotations. This gives a factorization of linear Gaussian stochastic maps in terms of rotations, scalings, drifts, and noise injections. Unlike Bloch--Messiah decompositions in continuous-variable quantum optics, this factorization does not use a symplectic structure.

\subsection{Gaussian Gate Library}\label{sec:gauss-lib}

\paragraph{\(\gname{PDisp}(\alpha)\)}
Shifts the mean by \(\alpha\): \(\mathcal N(\mu,\sigma^2)\mapsto\mathcal N(\mu+\alpha,\sigma^2)\). Triple \((1,\alpha,0)\).

\paragraph{\(\gname{PScale}(r)\)}
Scales mean and variance: \(\mathcal N(\mu,\sigma^2)\mapsto\mathcal N(e^r\mu,e^{2r}\sigma^2)\). Triple \((e^r,0,0)\). Reversible; no noise injection.

\paragraph{\(\gname{PMix}_{ij}(\theta)\)}
Planar rotation on two modes: \((X_i,X_j)\mapsto(\cos\theta\,X_i-\sin\theta\,X_j,\sin\theta\,X_i+\cos\theta\,X_j)\). Triple \((R(\theta),\bm 0,0)\) with \(R(\theta)\in\mathrm{SO}(2)\).

\paragraph{\(\gname{P2Sq}_{ij}(r)\)}
Hyperbolic rotation on two modes: \((X_i,X_j)\mapsto(\cosh r\,X_i+\sinh r\,X_j,\sinh r\,X_i+\cosh r\,X_j)\). Induces \(\Cov(X_i',X_j')=\sinh(2r)\) for independent \(\mathcal N(0,1)\) inputs.

Together with interferometer decompositions, these gates realize every element of \(\mathcal S_N\) with \(\Delta=0\) at depth \(O(N)\).

\subsection{Generator-derived gates: Fokker--Planck building blocks}

\paragraph{Backward vs forward generators.}
The Fokker--Planck (forward) generator \(\mathcal L^\dagger\) acts on densities; the backward (Kolmogorov) generator \(\mathcal L\) is its formal adjoint and acts on observables:
\begin{align}
    \mathcal L^\dagger p &= -\nabla\!\cdot\!\bigl[\bm\mu(x)\,p\bigr]+\tfrac12\sum_{ij}\partial_i\partial_j\bigl[D_{ij}(x)\,p\bigr],
    \\
    \mathcal L f &= \bm\mu(x)\!\cdot\!\nabla f+\tfrac12\sum_{ij}D_{ij}(x)\,\partial_i\partial_j f,
\end{align}
related by \(\langle f,\mathcal L^\dagger p\rangle=\langle\mathcal L f,p\rangle\). The finite-time propagator is \(P_t=\exp(t\mathcal L^\dagger)\) on densities and \(\exp(t\mathcal L)\) on observables; we report \(\mathcal L\) for each gate below. For linear drift and constant diffusion, \(\mathcal L^\dagger\) preserves the Gaussian manifold and the \(\mathcal S_N\)-triple of Eq.~\eqref{eq:gauss-propagator} is the propagator data.

\paragraph{Drift gate \(\gname{PDrift}(v,t)\).}
\(\mathcal L=v\,\partial_x\); SDE \(\dd X_t=v\,\dd t\); propagator \(p(x)\mapsto p(x-vt)\); triple \((1,vt,0)\). A generator-time-parametrized form of \(\gname{PDisp}\) suited to Trotter schedules.

\paragraph{Diffusion gate \(\gname{PDiff}(D,t)\).}
\(\mathcal L=D\,\partial_x^2\); SDE \(\dd X_t=\sqrt{2D}\,\dd W_t\); Gaussian convolution kernel \(K_t(y\mid x)=\mathcal N(y;x,2Dt)\); triple \((1,0,2Dt)\). No stationary distribution on \(\R\); the canonical noise-injection generator gate.

\paragraph{Ornstein--Uhlenbeck gate \(\gname{POU}(\gamma,D,t)\).}
\(\mathcal L=-\gamma\,x\,\partial_x+D\,\partial_x^2\); SDE \(\dd X_t=-\gamma X_t\,\dd t+\sqrt{2D}\,\dd W_t\); kernel
\begin{equation}
    K_t(y\mid x_0)=\mathcal N\!\Bigl(y;\;x_0e^{-\gamma t},\;\tfrac{D}{\gamma}\bigl(1-e^{-2\gamma t}\bigr)\Bigr);
\end{equation}
triple \(\bigl(e^{-\gamma t},0,\tfrac{D}{\gamma}(1-e^{-2\gamma t})\bigr)\); stationary distribution \(\mathcal N(0,D/\gamma)\). A non-trivial element of \(\mathcal S_1\) with both deterministic contraction (\(M=e^{-\gamma t}<1\)) and noise injection (\(\Delta>0\)); the pmode generator-derived example analogous to \(\gname{PIsing}\) for pbit dynamics.

\paragraph{Geometric Brownian motion gate \(\gname{PGBM}(\mu,\sigma,t)\).}
\(\mathcal L=\mu x\,\partial_x+\tfrac{\sigma^2 x^2}{2}\partial_x^2\) on \(x>0\); SDE \(\dd X_t=\mu X_t\,\dd t+\sigma X_t\,\dd W_t\) (multiplicative noise); closed form \(X_t=X_0\exp((\mu-\sigma^2/2)t+\sigma W_t)\). Generally not in \(\mathcal S_1\) due to state-dependent diffusion.

\paragraph{Cox--Ingersoll--Ross gate \(\gname{PCIR}(\kappa,\theta,\sigma,t)\).}
\(\mathcal L=\kappa(\theta-x)\,\partial_x+\tfrac{\sigma^2 x}{2}\partial_x^2\) on \(x\geq0\) with Feller condition \(2\kappa\theta\geq\sigma^2\); SDE \(\dd X_t=\kappa(\theta-X_t)\,\dd t+\sigma\sqrt{X_t}\,\dd W_t\); stationary distribution in the Gamma or non-central-\(\chi^2\) family.

\paragraph{Exchange generator.}
\(\mathcal L_{\mathrm{exch}}=\gamma(x_j-x_i)\partial_{x_i}+\gamma(x_i-x_j)\partial_{x_j}+D(\partial_{x_i}^2+\partial_{x_j}^2)\) drives two coupled OU processes conserving \(x_i+x_j\).

\paragraph{Correlated-noise generator.}
\(\mathcal L_{\mathrm{corr}}=D\partial_{x_i}^2+D\partial_{x_j}^2+2\rho D\partial_{x_i}\partial_{x_j}\) is pure noise injection with covariance \(D\bigl(\begin{smallmatrix}1&\rho\\\rho&1\end{smallmatrix}\bigr)\) per unit time.

\paragraph{Disjoint-support tensor identity.}
If \(\mathcal L_A,\mathcal L_B\) are Fokker--Planck operators with disjoint mode supports, then
\begin{equation}
    \exp\bigl[(\mathcal L_A\otimes\Id_B+\Id_A\otimes\mathcal L_B)\,t\bigr]
    =\exp(\mathcal L_A\,t)\otimes\exp(\mathcal L_B\,t),
    \label{eq:cv-disjoint-tensor}
\end{equation}
the continuous-variable analog of the disjoint-support identity used throughout Sections~\ref{sec:stoc_circs}--\ref{sec:trotterization} and the basis for parallel scheduling of pmode gates.

\paragraph{Hamiltonian--Langevin and BAOAB decomposition.}
Underdamped Langevin dynamics in \((\bm q,\bm p)\)-space with potential \(U(\bm q)\),
\begin{equation}
    \begin{aligned}
    \dd\bm q
    &=
    (\bm p/m)\,\dd t,
    \\
    \dd\bm p
    &=
    -\nabla U(\bm q)\,\dd t
    -\gamma\bm p\,\dd t
    +\sqrt{2\gamma m k_B T}\,\dd\bm W,
    \end{aligned}
\end{equation}
has Fokker--Planck generator \(\mathcal L=\mathcal L_{\mathrm{drift}}+\mathcal L_{\mathrm{fric}}+\mathcal L_{\mathrm{diff}}\) with
\begin{align}
    \mathcal L_{\mathrm{drift}}&=-(\bm p/m)\!\cdot\!\nabla_{\bm q}+\nabla U(\bm q)\!\cdot\!\nabla_{\bm p},
    \\
    \mathcal L_{\mathrm{fric}} &= \gamma\,\nabla_{\bm p}\!\cdot\!(\bm p\,\cdot),
    \\
    \mathcal L_{\mathrm{diff}} &= \gamma m k_B T\,\nabla_{\bm p}^2.
\end{align}
Each generator maps to a gate: \(\mathcal L_{\mathrm{fric}}\) gives a \(\gname{PScale}\) (momentum contraction), \(\mathcal L_{\mathrm{diff}}\) gives a \(\gname{PDiff}\) with \(D=\gamma m k_B T\), and \(\mathcal L_{\mathrm{drift}}\) gives a symplectic step. The BAOAB Strang integrator applies these in order: momentum half-step (B), position half-step (A), OU step (O), position half-step (A), momentum half-step (B). The sampler compiles to an alternating sequence of \(\gname{PDrift}\), \(\gname{PScale}\), and \(\gname{PDiff}\) gates, with Tier-3 \(\gname{CFunc}[-\nabla U]\) gates when \(U\) is non-quadratic.

\subsection{Tier hierarchy of non-Gaussian primitives}

We organize pmode constructions into tiers by the resources they consume: Tier 1 comprises the noiseless Gaussian gates of \ref{sec:gauss-lib}; Tier 2 adds prepared-and-discarded ancillas, continuous (2a) or discrete (2b); Tier 3 adds measurement and classical feedback.

\paragraph{Tier 2a: noise via auxiliary pmodes.}
For input \(X\sim\mathcal N(\mu,\sigma^2)\) and independent ancilla \(A\sim\mathcal N(0,\sigma_A^2)\), applying \(\gname{PMix}(\theta)\) and discarding \(A\) gives
\begin{equation}
    X'\sim\mathcal N\!\bigl(\cos\theta\cdot\mu,\;\cos^2\theta\,\sigma^2+\sin^2\theta\,\sigma_A^2\bigr).
\end{equation}
Correlated multi-mode noise is realized by correlating auxiliaries with \(\gname{P2Sq}(r)\) before mode-by-mode mixing; the resulting covariance \(\sin^2\theta\,\sigma_A^2\bigl(\begin{smallmatrix}\cosh 2r&\sinh 2r\\\sinh 2r&\cosh 2r\end{smallmatrix}\bigr)\) realizes any \(2\times2\) noise injection by tuning \((\theta,r,\sigma_A)\).

\paragraph{Tier 2b: Gaussian mixtures via auxiliary pdits.}
The controlled-shift gate is
\begin{equation}
    \begin{aligned}
    \gname{CShift}(\bm\alpha):\;
    \pket{k}\otimes p(x)
    &\longmapsto
    \pket{k}\otimes p(x-\alpha_k),
    \\
    \bm\alpha
    &=
    (\alpha_0,\ldots,\alpha_{d-1}).
    \end{aligned}
\end{equation}
Preparing a pdit with PMF \(\pi=(\pi_0,\ldots,\pi_{d-1})\), applying \(\gname{CShift}(\bm\alpha)\) to \((K,X)\) with \(X\sim\mathcal N(\mu,\sigma^2)\), and marginalizing \(K\) yields
\begin{equation}
    p_{X'}(x) \;=\; \sum_{k=0}^{d-1}\pi_k\,\mathcal N(x;\mu+\alpha_k,\sigma^2),
    \label{eq:cshift-mixture}
\end{equation}
a Gaussian mixture with shared variance. The companion controlled-scale gate \(\gname{CScale}(\bm r):\pket{k}\otimes p(x)\mapsto\pket{k}\otimes e^{-r_k}p(e^{-r_k}x)\) gives heteroscedastic mixtures \(\sum_k\pi_k\mathcal N(e^{r_k}\mu,e^{2r_k}\sigma^2)\). Finite Gaussian mixtures can approximate broad classes of continuous densities, including compactly supported densities in \(L^1\) under standard regularity assumptions. Tier-2b therefore gives a practical density-approximation route by adding one type of discrete ancilla and one controlled gate.

\paragraph{Tier 3: measurement-feedback for arbitrary nonlinearities.}
For state-dependent nonlinearities the measurement gate \(\gname{Meas}:\pmode\to\creal\) samples \(x_0\sim p(x)\) and produces a classical real; \(\gname{CFunc}[f]:\creal\to\creal\) evaluates an arbitrary measurable \(f\) on a classical coprocessor; and \(\gname{FeedbackDisp}:(\creal,\pmode)\to\pmode\) treats the classical value as a deterministic displacement on the live pmode. The composition \(\gname{Meas}\to\gname{CFunc}[\phi]\to\gname{Inject}\) realizes the pushforward \(\phi_*p_X\); inserting a \(\gname{PDiff}\) step recovers an overdamped Langevin update \(X_{t+\Delta t}=X_t-\Delta t\,\nabla V(X_t)+\sqrt{2\Delta t}\,\xi_t\) that samples from the Gibbs distribution \(\pi(x)\propto e^{-V(x)}\). Tier 3 incurs the round-trip latency cost of each measurement and classical evaluation; this compilation strategy is reserved for state-dependent nonlinearities that cannot be approximated by a finite mixture within the application accuracy budget.

\subsection{Hybrid pdit/pmode gates and jump-diffusion}

The general jump-diffusion process
\begin{equation}
    \begin{aligned}
    \dd X_t
    &=
    \mu_{N_t}(X_t)\,\dd t
    +
    \sigma_{N_t}(X_t)\,\dd W_t,
    \\
    N_t &: i\to j
    \text{ at rate }\lambda_{ji}(X_t),
    \end{aligned}
\end{equation}
has joint generator
\begin{equation}
    \mathcal L_{\mathrm{JD}}
    =\sum_i\pket{i}\!\pbra{i}\otimes\mathcal L_i^{(\mathrm{diff})}
    +\sum_{i\neq j}\lambda_{ji}(x)\bigl[\pket{j}\!\pbra{i}\otimes\mathcal K_{ji}-\pket{i}\!\pbra{i}\otimes\Id\bigr],
\end{equation}
with diffusion piece \(\mathcal L_i^{(\mathrm{diff})}=\mu_i(x)\partial_x+\tfrac{1}{2}\sigma_i^2(x)\partial_x^2\). Assuming constant rates and no reset, the Strang split
\begin{equation}
    P_{\Delta t}\;\approx\;e^{\mathcal L^{(\mathrm{diff})}\Delta t/2}\,e^{\mathcal L^{(\mathrm{jump})}\Delta t}\,e^{\mathcal L^{(\mathrm{diff})}\Delta t/2}
    \label{eq:jd-strang}
\end{equation}
has local error \(O(\Delta t^3)\) and global error \(O(\Delta t^2)\); the corresponding circuit alternates \(\gname{CtrlOU}\) (half-step), a pdit-only jump layer (full step), and \(\gname{CtrlOU}\) (half-step). State-dependent rates \(\lambda_{ji}(x)\) require a Tier-3 measurement of the pmode before each jump layer, and this hybrid Tier-2/Tier-3 schedule is a compilation for regime-switching dynamics with state-dependent rates that appear in financial, biophysical, and chemical models.

\section{X0 sampling circuits and the XTR-0 platform}
\label{app:x0-hardware}

\begin{figure}[ht]
\centering
\includegraphics[width=0.95\columnwidth]{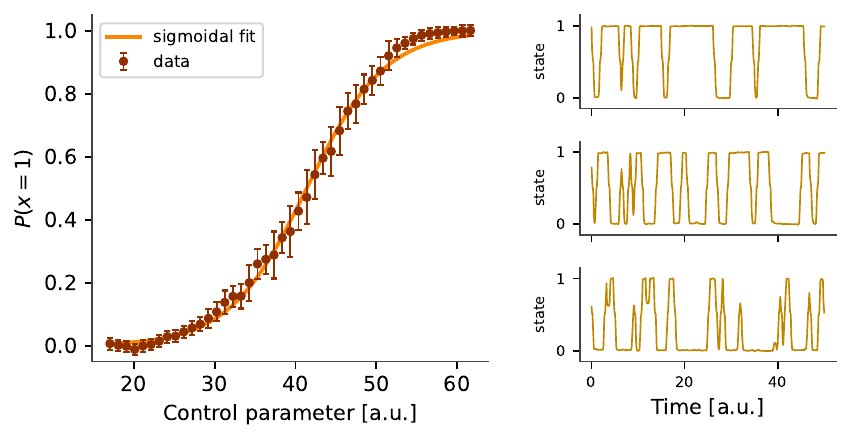}
\caption{Measured operating characteristic of an X0 pbit. The probability of observing state \(1\) is approximately sigmoidal in the control parameter (left); output waveforms at three bias settings show the corresponding telegraph signals (right). Data digitized from the published figure in the public X0/XTR-0 description.}
\label{fig:x0-pbit}
\end{figure}

This appendix summarizes the X0 test chip and the XTR-0 platform used to host it in the hardware demonstration of Section~\ref{sec:hw-experiment}, complementing the backend description of Section~\ref{sec:hard-disc}. The device physics of the X0 primitives is developed in Ref.~\cite{freitas2026taming}, and the platform is described in the documentation materials~\cite{extropic2025insidex0}. Here we collect the facts relevant to their use as a randomness source for PSCs.

\paragraph{Platform.}
XTR-0 is a desktop development platform consisting of a digital host, an FPGA, and two daughterboard sockets, each of which can receive a chip carrying probabilistic primitives. The socketed construction allows the same platform to host future chips without modification. The test chip currently hosted on the platform, the X0 chip, was fabricated in an advanced FinFET CMOS process~\cite{freitas2026taming} with the goal of validating transistor-level noise models and probabilistic circuit designs.  Control parameters are set through on-chip 8-bit digital-to-analog converters, while the output signals, which are weak because the circuits operate in the subthreshold regime, are routed off-chip through a dedicated amplification chain and digitized on the daughterboard before reaching the host~\cite{freitas2026taming}. In this way, the host observes the analog dynamics of every primitive directly and can update parameters in a closed loop.

\paragraph{Sampling model.}
Each primitive outputs a continuous-time random voltage signal whose stationary statistics realize a target distribution, such that observing the signal at intervals long compared to the relaxation time of the circuit yields approximately independent draws. For the pbit, the memory between observations separated by a lag \(\tau\) is quantified by the normalized autocorrelation
\begin{equation}\label{eq:pbit-autocorr}
    r_{xx}(\tau) \approx e^{-\tau/\tau_0},
\end{equation}
with \(\tau_0\approx 100\;\mathrm{ns}\) for the device used here~\cite{jelinvcivc2025efficient}, while pbit designs on the X0 process span relaxation times from roughly \(1\;\mathrm{ns}\) to \(1\;\mathrm{ms}\). For a representative Gaussian primitive, Ref.~\cite{freitas2026taming} reports correlation times down to \(\tau\approx 1\;\mathrm{ns}\) and an energy per independent sample, defined as the product of the DC power and the correlation time, of \(E\approx 15\;\mathrm{aJ}\). These figures anchor the energy and rate ranges quoted in Section~\ref{sec:hard-disc}. It is important to note that extreme values of the controls recover deterministic behavior: for example, the pbit becomes a constant 0 or 1 source.

\begin{table}[h]
\centering
\caption{X0 sampling circuits exposed by XTR-0.}
\label{tab:x0-circuits}
\resizebox{0.95\columnwidth}{!}{\begin{tabular}{llll}
\toprule
Circuit & Output signal & Distribution & Controls \\
\midrule
pbit & 2-level telegraph & Bernoulli & \(1\) (bias) \\
pdit & \(k\)-level telegraph & categorical & \(k-1\) (weights) \\
pmode & continuous & Gaussian & \(\sigma\) (1D); \(\rho\) (2D) \\
pMoG & multi-level & Gaussian mixture & \(\pi_j,\mu_j,\Sigma_j\) (bounded) \\
\bottomrule
\end{tabular}}
\end{table}

\paragraph{Probabilistic primitives.}
The X0 primitives supply hardware draws for the wire types of Table~\ref{tab:wire-types}, and are summarized in Table~\ref{tab:x0-circuits}. The pbit outputs a two-level random telegraph signal whose bias is approximately sigmoidal in a single control voltage, as shown by the measurements of Figure~\ref{fig:x0-pbit}. The pdit generalizes this behavior to a \(k\)-level signal, where normalization leaves \(k-1\) independent controls. Furthermore, XTR-0 exposes a one-dimensional pmode with programmable standard deviation and a two-dimensional pmode with programmable correlation between its two output voltages. The pmode electronic circuit is a network of capacitively coupled CMOS inverters, referred to in Ref.~\cite{freitas2026taming} as a non-equilibrium adjustable-temperature resistor network (NEAT-RN) and identified there as the first electronic implementation of a Brownian gyrator using CMOS circuits. The pmode wire of Section~\ref{sec:notation} is the software abstraction of this device. Finally, the mixture-of-Gaussians circuit (pMoG) jumps between levels whose positions, spreads, and weights are programmable within bounds.

\end{document}